\documentclass{pasj00}
\newcommand{\logn}{$\log{N}$-- $\log{S}$}
\newcommand{\ergs}{erg~cm$^{-2}$~s$^{-1}$}
\newcommand{\ergss}{erg~cm$^{-2}$~s$^{-1}$~sr$^{-1}$}
\newcommand{\photonss}{photon~keV$^{-1}$~cm$^{-2}$~s$^{-1}$~sr$^{-1}$}

\SetRunningHead{A.~Kushino et al.}{
X-Ray Background: Spectrum and its Large-Scale
Fluctuation
}
\Received{2002/01/07}%{yyyy/mm/dd}
\Accepted{2002/04/09}%{yyyy/mm/dd}

\title{
Study of the X-Ray Background Spectrum and its Large-Scale
Fluctuation with ASCA
}

\ifnum1=0
\author{
   Akihiro \textsc{Kushino}, 
   Yoshitaka \textsc{Ishisaki},\\ 
   Umeyo \textsc{Morita}, 
   Noriko Y.~\textsc{Yamasaki}, 
   Manabu \textsc{Ishida}, 
   Takaya \textsc{Ohashi} 
   }
\affil{
   Department of Physics, Tokyo Metropolitan University, 
   1-1 Minami-Osawa, Hachioji, \\ Tokyo 192-0397, Japan
   }
   \email{kushino@phys.metro-u.ac.jp}
   \and
\author{
   Yoshihiro \textsc{Ueda}
   }
\affil{
   The Institute of Space and Astronautical Science 
   (ISAS), 3-1-1 Yoshinodai, Sagamihara, \\ Kanagawa  229-8510, Japan
   }

\fi

\author{
   Akihiro \textsc{Kushino},\altaffilmark{1} 
   Yoshitaka \textsc{Ishisaki},\altaffilmark{2} \\
   Umeyo \textsc{Morita},\altaffilmark{2} 
   Noriko Y.~\textsc{Yamasaki},\altaffilmark{2} 
   Manabu \textsc{Ishida},\altaffilmark{2} 
   Takaya \textsc{Ohashi},\altaffilmark{2} 
   and Yoshihiro \textsc{Ueda}\altaffilmark{3}} 

\altaffiltext{1}
{National Space Development Agency of Japan (NASDA), 
2-1-1 Sengen, Tsukuba, Ibaraki 305-8505}
\email{kushino.akihiro@nasda.go.jp}
\altaffiltext{2}
{Department of Physics, Tokyo Metropolitan University, 1-1 Minami-Osawa, 
Hachioji, Tokyo 192-0397}
\altaffiltext{3}
{The Institute of Space and Astronautical Science (ISAS), 3-1-1 
Yoshinodai, Sagamihara, Kanagawa 229-8510}

\begin{document}
\maketitle

\begin{abstract}

We studied the energy spectrum and the large-scale fluctuation of the
X-ray background with the {\it ASCA} GIS instrument based on
the {\it ASCA} Medium Sensitivity Survey and Large Sky Survey observations.
A total of 91 fields with Galactic latitude $|b|>10^\circ$ were selected 
with a sky coverage of 50 deg$^2$ and 4.2 Ms of exposure.
For each field, non X-ray events were carefully subtracted and
sources brighter than $\sim 2\times 10^{-13}$ \ergs\ (2--10 keV)
were eliminated. Spectral fits with a single power-law model
for the individual 0.7--10 keV spectra showed a significant excess
below $\sim 2$ keV, which could be expressed by an additional thermal
model with $kT\simeq 0.4$ keV or a steep power-law model with a photon
index of $\Gamma^{\rm soft}\simeq 6$. The 0.5--2 keV intensities of
the soft thermal component varied significantly from field to field
by 1~$\sigma= 52^{+ 4}_{- 5}$~\%, and showed a maximum toward 
the Galactic Center. This component is considered to be entirely Galactic.
As for the hard power-law component, an average photon index of 91 fields
was obtained to be $\Gamma^{\rm hard} = 1.412\pm 0.007\pm 0.025$
and the average 2--10 keV intensity was calculated as
$F_{\rm X}^{\rm hard} = (6.38\pm 0.04\pm 0.64)\times 10^{-8}$
\ergss\ (1~$\sigma$ statistical and systematic errors).
The Galactic component is marginally detected in the hard band.
The 2--10 keV intensities show a 1~$\sigma$ deviation of
$6.49^{+ 0.56}_{- 0.61}~\%$, while deviation due to
the reproducibility of the particle background is 3.2\%.
The observed deviation can be explained by the Poisson noise
of the source count in the f.o.v.\ ($\sim 0.5$ deg$^2$),
even assuming a single \logn\ relation on the whole sky.
Based on the observed fluctuation and the absolute intensity,
an acceptable region of the \logn\ relation was derived,
showing a consistent feature with the recent {\it Chandra} and
{\it XMM-Newton} results.
The fluctuation of the spectral index was also examined; 
 it implied a large amount of hard sources and
a substantial variation in the intrinsic source spectra
($\Gamma_{\rm S}\simeq 1.1\pm 1.0$).
%% According to the recent {\it Chandra} results reported by
%% \citet{Rosati2002}, the X-ray background in 2--10 keV
%% has been resolved into discrete sources by 73--96~\%
%% at a flux limit of $S\gtsim 4.5\times 10^{-16}$ \ergs. 
\end{abstract}

\section{Introduction}
\label{sec:intro}

The Cosmic X-ray Background, hereafter called CXB, dominates the 
hard X-ray surface brightness (\cite{Fabian1992}), and was discovered
with the Aerobee sounding rocket carrying three Geiger counters
\citep{Giacconi1962} along with the first extra-solar X-ray source, Sco X-1.
The effort of resolving the CXB into discrete sources has produced a 
\logn\ relation, which gives the number density ($N$) of discrete
sources above a certain flux level ($S$).
This study made a major advance recently with the
advent of the {\it Chandra} X-ray Observatory.
The detection limit of {\it Chandra} is down to
$S\sim 10^{-16}$ \ergs\ in the 2--10 keV band,
and some of the sources are too faint for optical identifications to be 
made. In this energy range, more than half (60--90\%) of the CXB
emission has been resolved into discrete sources
(e.g.\ \cite{Mushotzky2000} for the SSA13 field;
\cite{Brandt2001} for the {\it Chandra} Deep Field North;
\cite{Tozzi2001} for the {\it Chandra} Deep Field South).
Moreover, recent {\it Chandra} and {\it XMM-Newton} results indicate a
turn-over of around $S\sim 2 \times 10^{-14}$ \ergs\ \citep{Campana2001,
Hasinger2001, Baldi2002}. Above this flux level,
the relation seems to be consistent with a uniform distribution of
sources in Euclidean space, namely, $N(>S)\propto S^{-1.5}$
%% \citep{Piccinotti1982,Hayashida1992,Ueda1999-LSS,Ueda1999}.
 (Piccinotti et al.\ 1982; Hayashida et al.\ 1992; Ueda et al.\ 1999a,b).
At the present time, the 2--10 keV CXB has been almost resolved
into discrete sources, leaving at most 10--20\% at the faintest
flux limit. The main uncertainty of the resolved fraction
lies in the measurement of the total intensity of the CXB,
which requires a well-calibrated detector with a low internal background,
as well as large sky coverage.

It also became clear that the emission consists of roughly
two components. The hard-band emission in $\sim 2$--10 keV shows
almost an isotropic distribution in the sky, and the spectrum is
characterized by a power-law spectrum with a photon index of  
$\Gamma\simeq 1.4$ (e.g.\ \cite{Marshall1980}: 3--15 keV;
\cite{Gendreau1995}: 1--10 keV; \cite{Miyaji1998}: 1--10 keV;
\cite{Vecchi1999}: 1--8 keV)\@.
The survey observations by {\it Uhuru} (2--7 keV)  
and {\it HEAO~1}\/ A-2 ($\sim 2$--10 keV)  
showed that the dipole amplitude after removing the Galactic 
contribution were $0.61\pm 0.26$\% (\cite{Protheroe1980}) and 
$\sim 0.5$\% (\cite{Shafer1983}), respectively.
Below $\sim 2$ keV, the spectrum becomes steeper with
$\Gamma\simeq 2.1$ (e.g.\ \cite{Hasinger1992} at $\sim 1$ keV)\@.
The {\it ROSAT} all-sky survey observation showed
significant structures in the soft X-ray background
which are correlated mainly with the Galactic distribution of the hot
gas \citep{Snowden1995}. As for the absolute intensity of the CXB,
there is fairly large uncertainty among the measurements; namely,
the reported intensities at 1 keV are $13.4\pm0.3$ \citep{Hasinger1992},
$9.4\pm 0.4$ \citep{Gendreau1995} and
$10.4^{+1.4}_{-1.1}$ \citep{Parmar1999} \photonss.
\citet{Barcons2000} compiled previous measurements of the CXB intensity
with the {\it ASCA}, {\it BeppoSAX} and {\it ROSAT}
and pointed out that these differences are primarily
caused by the cosmic variance, i.e.\ spatial variation
of source count due to the limited solid angle of sky coverage,
and in some part resulted from instrumental cross-calibration errors
and subtraction process of the Galactic contribution.

X-ray surveys with optical follow-up observations
(e.g. \cite{Schmidt1998}; \cite{Akiyama2000}; \cite{Ishisaki2001})
have identified many active galactic nuclei (AGNs)
and provided information about their spectral evolution.
There is a significant discrepancy between the spectral
indices of the CXB above $\sim 2$ keV ($\Gamma\simeq 1.4$) and of
type-1 Seyfert galaxies ($\Gamma\simeq 1.7$), which led to a model
that a significant fraction of AGN is heavily absorbed
(\cite{Madau1994}; \cite{Comastri1995}; \cite{Gilli2001}).
In fact, for the sources with $S\gtsim 10^{-15}$ \ergs,
the {\it XMM-Newton} satellite revealed that around
half of them are strongly absorbed in their spectra \citep{Hasinger2001}.
In the flux range $S\sim 10^{-14}$--$10^{-11}$ \ergs,
a spectral survey from {\it ASCA} showed that the average
spectral slope becomes progressively harder as the sources become fainter,
and the observed photon index of 2.1 at around $10^{-11}$ \ergs\ turns to
1.6 at around $10^{-13}$ \ergs\ in the 0.7--10 keV band \citep{Ueda1999}.
A precise determination of the spectral shape of the CXB
and an investigation of its field-to-field difference
would provide rich information to understand
the spectral evolution of AGNs taking place around the flux of
$10^{-13}$ \ergs.

The {\it ASCA} Medium-Sensitivity Survey (AMSS) is a serendipitous
survey, consisting of 368 fields which can be regarded as a
random sampling of the CXB\@.
\citet{Ueda2001} produced a catalog of 1343 X-ray sources
based on the AMSS, and gave a fairly tight \logn\ relation
in the flux range above $7 \times 10^{-14}$ \ergs\ \citep{Ueda1999}.
Also, combined with the {\it ASCA} Large Sky Survey
%%\citep{Ueda1998,Ueda1999-LSS},
(Ueda et al.\ 1998, 1999a),
the main aim of the present study is to determine
the absolute intensity and the spectral shape of the CXB
as precisely as possible.
The data after the point-source elimination formed the basis
for the present study.
It would also enable us to constrain the \logn\ relation
or the intrinsic spectral distribution of the constituent sources
through the fluctuation analysis,
as well as to look into the large-scale distribution of the CXB\@.
The GIS system \citep{Ohashi1996, Makishima1996} consists of
two gas imaging spectrometers,
which have well-studied and low internal background, 
simultaneously providing a large ($\gtrsim 40'$ in diameter)
field of view (f.o.v.), when compared with the {\it XMM-Newton}
PN and MOS CCDs and {\it Chandra} ACIS CCDs.
These properties are powerful for studying the spectral features of the CXB\@.

The remainder of this paper is organized as follows.
In section \ref{sec:Obs}, we discuss the selection criteria of the fields
employed in the present paper for the CXB analysis and explain how 
energy spectra are extracted.
In section \ref{sec:Res}, we present the fitting results.
In section \ref{sec:Large}, the large-scale distributions
in the spectral parameters are investigated.
In section \ref{sec:logNlogS},
using simulations which take into account the complicated response of
{\it ASCA}, we evaluate the CXB intensity
and constrain the \logn\ relation and the spectral distribution of sources
from the observed fluctuation.
In section \ref{sec:Dis}, we discuss implications of the results,
and the summary will be given in section \ref{sec:Con}.

\section{Observation and Analysis}
\label{sec:Obs}
The present study was carried out with a GIS instrument consisting 
 of GIS~2 and GIS~3 sensors (\cite{Ohashi1996}; \cite{Makishima1996}).
The detectors had a wide ($\sim 40'$ in diameter) f.o.v.\ and
a moderate spatial resolution of $\sim 3'$ (Half Power Diameter) 
combined with the X-Ray Telescope (XRT; \cite{Serlemitsos1995}).
Our primary goal of the analysis was to determine
the parameters of the CXB for each direction
which cannot be resolved into discrete sources with the GIS,
namely, the {\em unresolved} CXB\@.
Due to the low surface brightness of the unresolved CXB,
we had to take into account several issues that would not matter
very much in point-source studies.
First of all, the sample fields must be carefully selected
in order to obtain reasonable constraints on the parameters.
The second one is to minimize and precisely estimate
the amount of fake signals caused mainly by charged particles,
which we call the Non X-ray Background (NXB)\@.
The third one is to eliminate {\it resolved} sources
in the GIS f.o.v.
The last one is the instrumental response for the unresolved CXB\@.

\subsection{Selection of the Fields}
\label{subsec:selection}
Our sample fields were primarily selected from the first AMSS fields
(\cite{Ueda2001}), which consisted of 368 combined fields observed
between 1993 May and 1996 December 
with the Galactic latitude $|b| > 10^\circ$.
We also included the field of the {\it ASCA} Large Sky Survey
%% (LSS; \cite{Ueda1998}; \cite{Ueda1999-LSS}), 
(LSS; Ueda et al.\ 1998, 1999a),  
dividing the LSS 
field into four,
labeling them a-LSS, b-LSS, c-LSS, and d-LSS, from south-west to north-east.

In our prompt sample,
several fields were composed of multiple pointing observations,
covering a larger sky area than that of a single pointing.
Furthermore, most of the AMSS fields were pointed to certain
targets and often contained rather bright X-ray sources to study the CXB\@.
Since the Point Spread Function (PSF) of the XRT has long outskirt, 
approximated by 
\begin{equation}\label{eq:PSF}
{\rm PSF}(r) = \exp\left(-{r\over 1'.02}\right)
       + 0.0094\; \exp\left(-{r\over 6'.14}\right),
\end{equation}
a bright target source would contaminate the whole detector area ($r\ltsim 20'$).
We therefore selected fields based on the following criteria:
\begin{enumerate}
\item  Total exposure time summed up over all pointings is longer than
$\sim 20$ ks.
\item  No known extended X-ray sources is present in the field,
e.g., nearby galaxies, star-forming regions, or supernova remnants.
\item  The mean count rate including the NXB is lower 
than 0.3 count~s$^{-1}$ per one GIS sensor in 0.7--10 keV,
accumulated within a radius of 22 mm ($\simeq 22'$)
from the optical axis of each sensor.
For multi-pointing observations, we accept fields when at least
one pointing fulfills this condition.
\item  The remained sky area after the source elimination
(``Area~2'' column of table~\ref{tab:obs}) is more than
two thirds of the original sky area (``Area~1'').
\end{enumerate}
Exactly 100 fields satisfied these conditions.

\begin{figure}[htb] 
\begin{center}
\FigureFile(0.24\textwidth,0cm){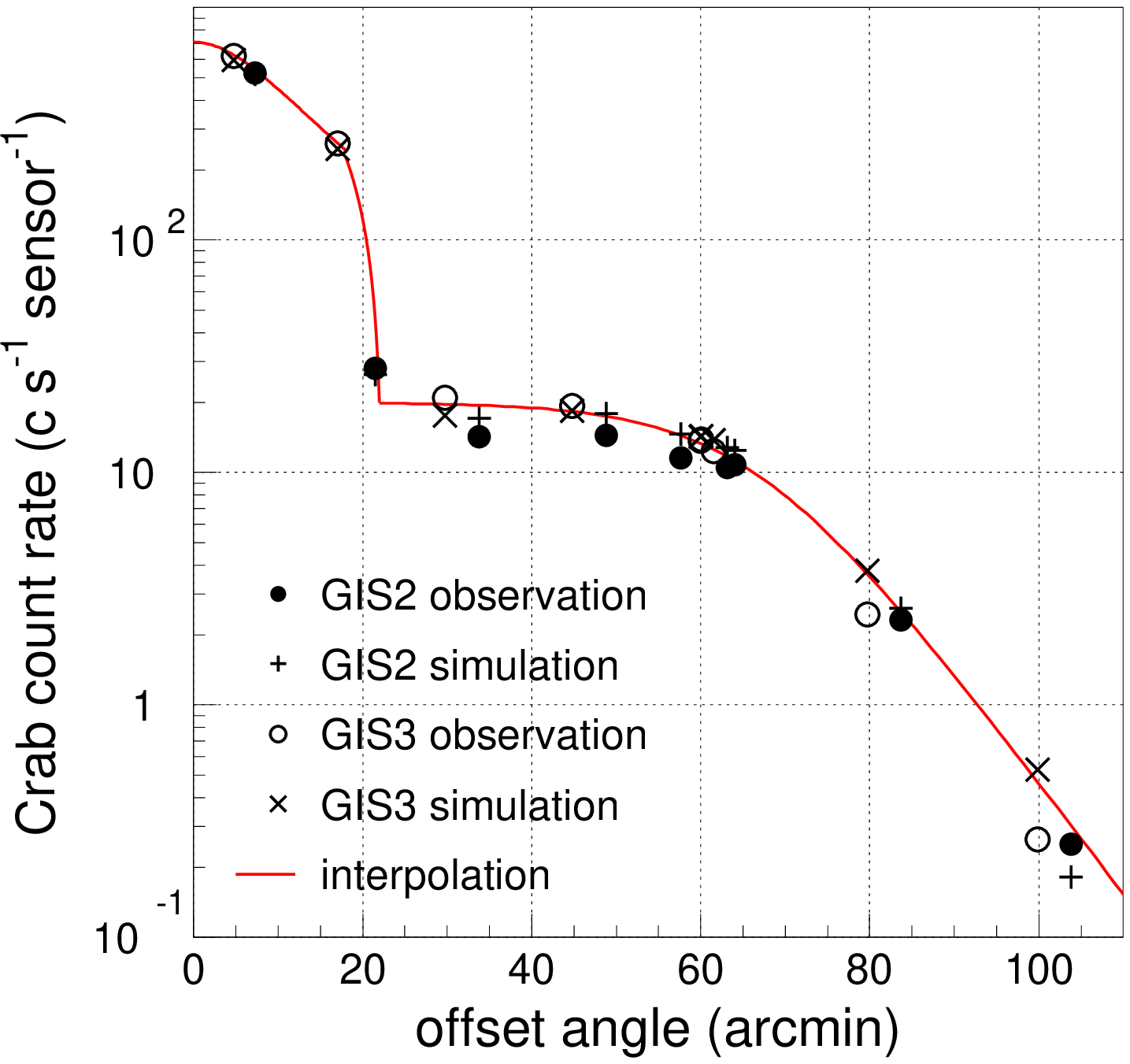}
\FigureFile(0.24\textwidth,0cm){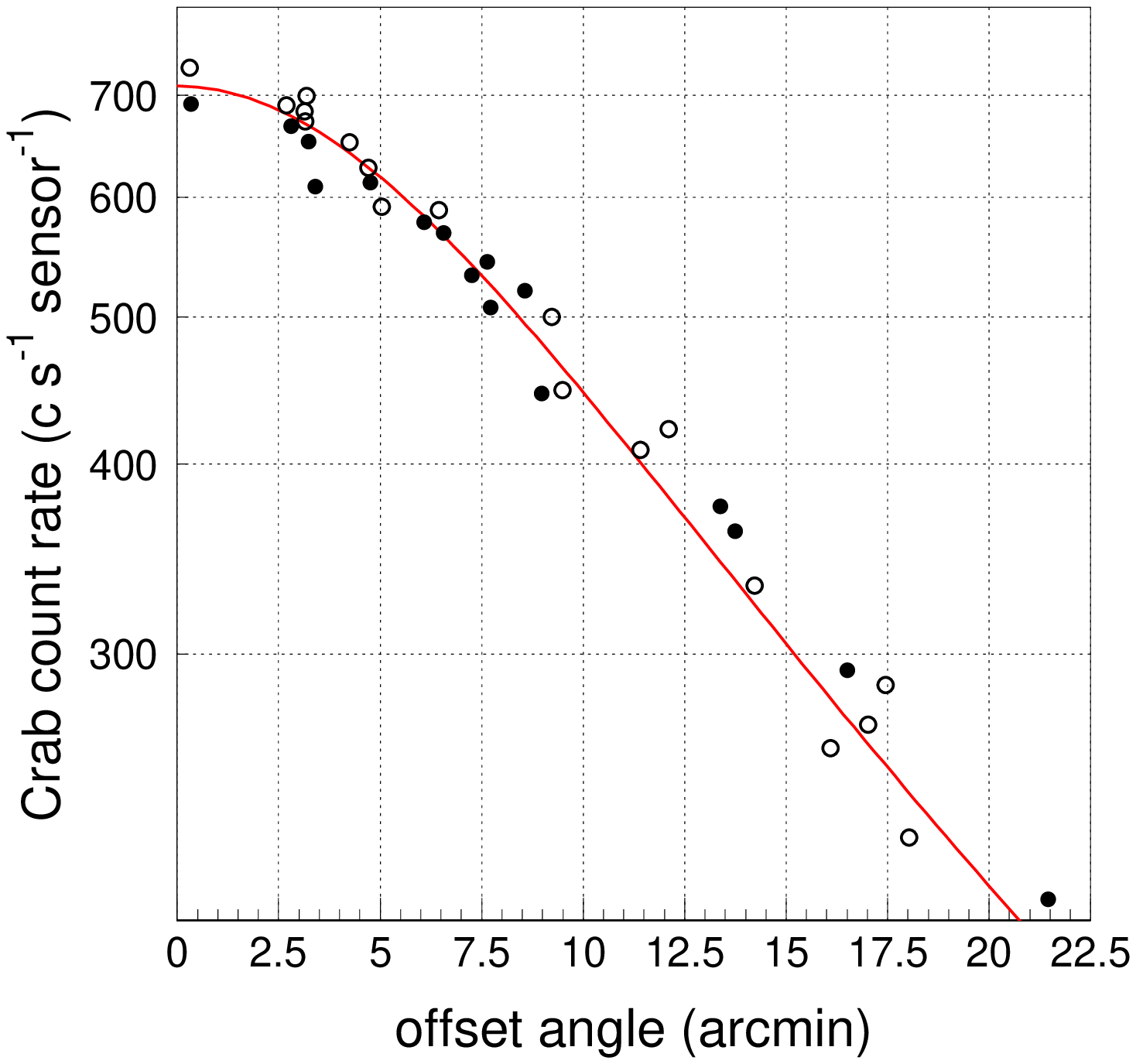}
\end{center}
\caption{0.7--10 keV count rate of the Crab nebula observed with GIS
at various offset angles from the optical axis of each GIS sensor.
Events within a radius of 20 mm ($\simeq 20'$) were accumulated,
and the CXB and the NXB counts were subtracted
after being corrected for the dead time.
The left panel is shown for a wide range of offset angles of up to $\sim 
100$ arcmin.
The observed count rates for each sensor are plotted in circles,
as well as the expectations from the simulation plotted in crosses.
The solid line represents an interpolated function using analytical formulae.
The right hand panel is a close-up, when the Crab nebula is placed
inside the GIS f.o.v.
}\label{fig:Crab}
\end{figure}

Since the XRT allows some fraction of photons from outside of the f.o.v.\ to
pass through (stray light; \cite{Tsusaka1995}),
it is a significant problem in the CXB study with {\it ASCA}\@.
Figure~\ref{fig:Crab} represents the 0.7--10 keV count rate of the Crab nebula
pointed at various offset angles from the optical axis of each GIS sensor.
When the Crab nebula was placed at $\sim 1^\circ$ away from the optical axis,
namely far outside of the GIS f.o.v., its count rate was still larger than
10 count~s$^{-1}$~sensor$^{-1}$ in 0.7--10 keV\@.
Since the CXB gave only $\sim 0.12$ count~s$^{-1}$~sensor$^{-1}$
in the same energy band, the stray flux needed to be carefully examined.
In order to avoid any influence of the stray light in the
present analysis, we further selected those fields which had no bright
sources around the f.o.v.
We mainly consulted with the {\it ROSAT} All-Sky Survey Bright Source Catalog
(RASS-BSC; \cite{Voges1999}), which lists sources with a flux limit of
$2.8\times 10^{-13}$ \ergs, assuming a power-law photon index of $\Gamma=2$
in the 0.1--2.4 keV energy band.
However, the sensitive energy range of {\it ROSAT} is
limited to below $\sim 2$ keV and sources with hard
spectra or strong absorption would be missed.
The {\it HEAO~1}\/ A-1 X-ray source catalog (\cite{Wood1984})
was also used to follow up such cases.
{\it HEAO~1}\/ A-1 is sensitive above $\sim 2$ keV,
and the flux limit is $4.78\times 10^{-12}$ \ergs\ in 2--10 keV
for a Crab-like spectrum.

\begin{figure}[htb] 
\begin{center}
\FigureFile(0.5\textwidth,0.5\textwidth){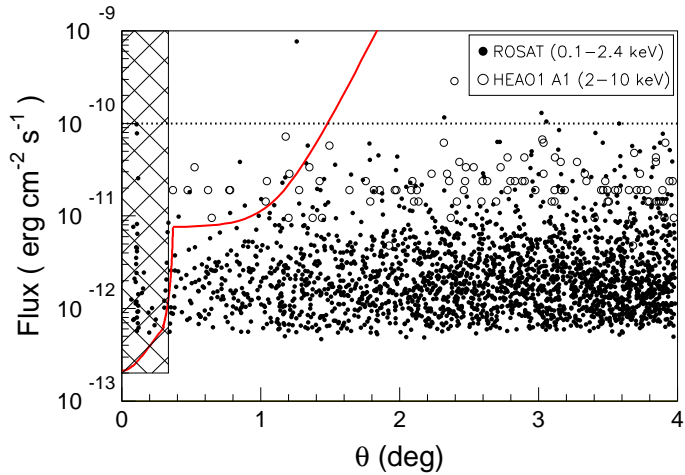}
\end{center}
\caption{Fluxes of cataloged bright sources within 4$^\circ$ from the
center of 100 pointed fields. The filled and open circles correspond to the
sources in the RASS-BSC
(\cite{Voges1999}) and in the {\it HEAO~1}\/ A-1 X-ray source catalog
(\cite{Wood1984}), respectively. The dotted line is used to discriminate
the {\it HEAO~1} sources which have large position errors, and the solid
curve is for the {\it ROSAT} sources. Sources lying above these lines
and outside of the GIS field (hatched region) can affect the measured
CXB flux by more than 2.5\%. And the relevant fields were discarded
from the analysis.
}\label{fig:rass+a1-cut}
\end{figure}

Based on these catalogs, the distance distribution of the cataloged sources
from the centers of the selected 100 fields was determined as shown in
figure~\ref{fig:rass+a1-cut}.
In this figure, the source intensities are plotted
as a function of the offset angle.
The solid curve indicates a level where
sources above it could contribute to the CXB intensity
accumulated within the whole GIS f.o.v.\ ($r < 20$ mm $\simeq 20'$)
by more than 2.5\%.
This estimation is based on a calibration
observation of the Crab nebula at various offset angles,
which is indicated in figure~\ref{fig:Crab} by the solid line.
Using the {\it ROSAT} sources,
we excluded 9 fields in which there were sources above the 2.5\% line
{\em outside} the GIS f.o.v.\ ($r > 20$ mm $\simeq 20'$).
We did not exclude fields if bright sources are {\em inside} the f.o.v.,
i.e.\ in the hatched region in figure~\ref{fig:rass+a1-cut},
since they would be eliminated in the source-elimination process
(see subsection \ref{subsec:elimination}).
As for the {\it HEAO~1} sources, any fields were excluded
if sources brighter than $1.0\times 10^{-10}$ \ergs\ were present
within $r<4^\circ$. This is because the positional errors of the
{\it HEAO~1} sources are quite large ($\sim 1^\circ$).
One field was picked up by the {\it HEAO~1}-source selection;
 however, it had already been marked by the {\it ROSAT}-source selection.
We therefore excluded 9 fields in total.
One of the LSS fields (a-LSS) was rejected here.

\begin{figure}[htb] 
\begin{center}
\FigureFile(0.5\textwidth,0.3\textwidth){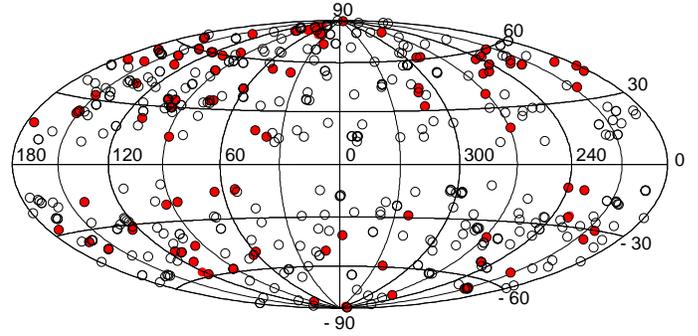}
\end{center}
\caption{Distribution of the 91 analyzed fields, shown with filled
circles, in the Galactic coordinate.  The open circles are other AMSS
 and LSS fields. 
The plotted symbols do not indicate the actual size of the
observed field.
}\label{fig:FIELDS}
\end{figure}

After these field selections, we ended up with 91 fields
whose sky distribution is shown in figure~\ref{fig:FIELDS}
together with the original AMSS and LSS fields.
The names and coordinates of the fields are listed in table~\ref{tab:obs}.
Among them, 33 fields consist of multi-pointing observations.
The exposure time of the 91 (88) fields amounts to 4240.54 (3844.27) ks and
the covering sky area extends 49.53 (42.77) deg$^2$,
where the numbers in parentheses denote those without the three LSS fields.
According to the AMSS and the LSS catalogs,
the brightest source in the 91 fields is
1AXGJ 085449+2006 with 0.7--7 keV flux of
$5.89\times 10^{-12}$ \ergs\ in OJ 287 field,
or 1AXG J004847+3157 with 2--10 keV flux of
$7.58\times 10^{-12}$ \ergs\ in Mrk 348 field,
depending on the chosen energy band.
Both sources correspond to the target, itself, i.e.\ OJ 287 or Mrk 348.
Among the 88 AMSS fields,
there are 184 (141) sources with their 0.7--7 (2--10) keV flux brighter than
$2\times 10^{-13}$ \ergs\ in each energy band in the AMSS catalog.

\subsection{Data Screening and NXB Subtraction}
\label{subsec:NXB}
We screened all of the GIS events detected in each sample field,
first employing the standard event selection criteria:
(1) the GIS should be in the nominal observation mode,
i.e.\ the PH normal mode with the nominal bit assignment,
the spread discriminator turned on,
and the satellite not in the South Atlantic Anomaly (SAA)
where the high-voltage supply of the GIS is switched off;
(2) the elevation of f.o.v.\ should be
$\geq 5^\circ$ (or $25^\circ$) above the night (or sunlit) Earth rim;
(3) the geomagnetic cut-off rigidity (COR) should be $\gtsim 6$ GV\@.
We performed further event screening utilizing the GIS monitor counts
(H0 + H2; see \cite{Ohashi1996}; \cite{Makishima1996}),
Radiation Belt Monitor (RBM) count, and COR\@.
This additional selection improved the reproducibility of the NXB estimation,
 mainly by rejecting data affected by sporadic increases in the NXB count,
which were due probably to the concentration of charged particles
on the satellite orbit.
A stricter condition for the RBM count was applied around the SAA and Hawaii.
Details of the screening procedure are described in \citet{Ishisaki1996}
and \citet{Ishisaki1997}.

The obtained event list contains not only the CXB photons, 
but also the NXB events, which must be carefully subtracted.
The corresponding spectra or images of the NXB were separately created,
using observations when the XRT was pointed to the night Earth.
Under the same condition as described in the previous paragraph,
we processed seven years of night-Earth data from 1993 June to 2000 May,
with a net exposure time of 4880.43 ks.
We further sorted both the observed data and the night-Earth data
into six intervals of the H0 + H2 monitor count in 5 count~s$^{-1}$ steps,
and performed NXB subtraction for each interval.
This is because the H0 + H2 monitor count,
which is tightly correlated with the NXB count (\cite{Makishima1996}),
can be used as a good NXB indicator during on-source observations.

\begin{figure}[htb] 
\begin{center}
\FigureFile(0.5\textwidth,0.5\textwidth){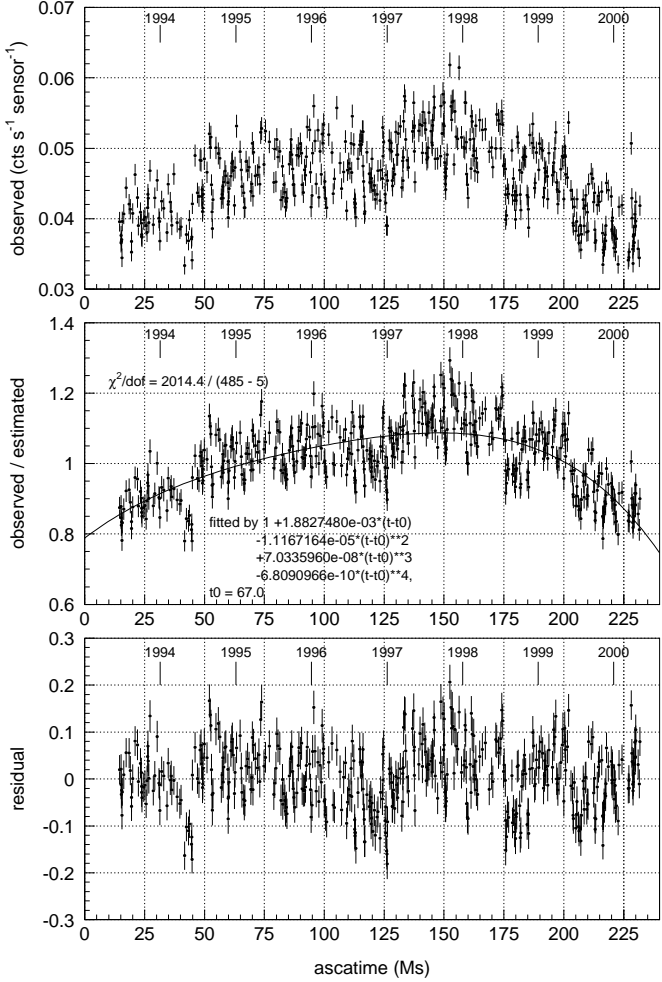}
\end{center}
\caption{
Top panel: Long-term change of the observed NXB count rate
(count~s$^{-1}$~sensor$^{-1}$) in the 0.7--7 keV band accumulated
within 20 mm ($\simeq 20'$) from the optical axis of each GIS sensor.
Each data point was determined from every 10 ks exposure of the night Earth
between 1993 June and 2000 May, and plotted as a function of
the mean {\it ASCA} time, which is defined as the elapsed time in second
since 1993 January 1.
Middle panel: observed NXB count rate divided by
the estimated one using the H0 + H2 monitor count.
The long-term trend is fitted by the fourth order polynomial,
which is indicated by the solid line.
Bottom panel: residuals of the fitting.
}\label{fig:NXB}
\end{figure}

There is also a long-term change of the NXB count rate, 
which is shown in figure~\ref{fig:NXB}.
The count rate gradually increased for the first four years,
and then turned to decreasing, peaking at around 1997--1998.
This long-term change was presumably caused by
a gradual drop in the satellite altitude and
the cycle of the solar activities with a period of $\sim 11$ yrs.
During the quiescent state of the solar activities,
which is at the minimum around 1996--1997,
the atmosphere shrunk so that the density of charged particles increased.
Since the observation date of our sample fields ranged
from 1993 June to 1996 December,
there could be about a $\pm 10$\% error in the NXB estimation
unless the long-term trend were corrected.
We fitted the trend by a fourth-order polynomial,
which was used for the correction.

\begin{figure}[htb] 
\begin{center}
\FigureFile(0.5\textwidth,0.5\textwidth){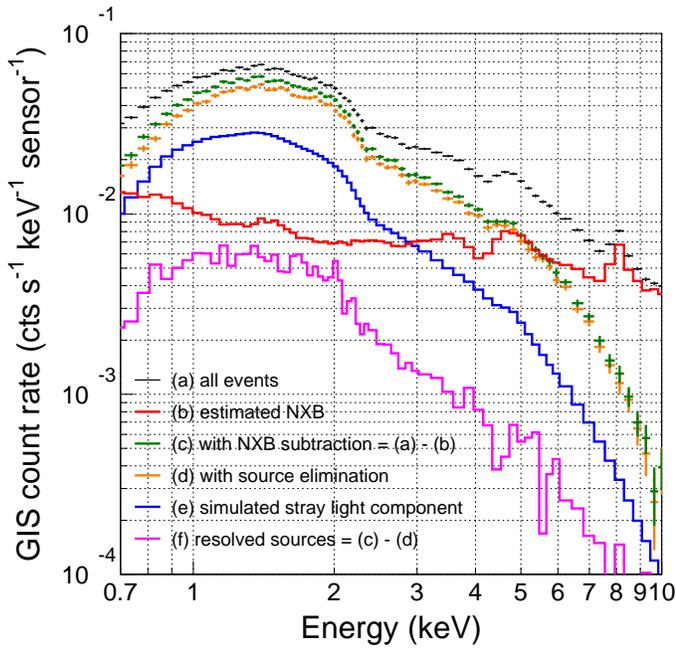}
\end{center}
\caption{
Comparison of the spectral components for the LSS field
when accumulated within a radius of 20 mm ($\simeq 20'$)
from the optical axis of each GIS sensor.
The spectra of all the pointings and both GIS sensors are summed up.
From upper to lower (at 8 keV), each spectrum represents:
(a) all events obtained in the LSS field after
the data screening described in subsection \ref{subsec:NXB};
(b) the NXB spectrum estimated in the way as described in subsection 
\ref{subsec:NXB};
(c) the LSS spectrum after the NXB subtraction, i.e.\ $\rm (a) - (b)$;
(d) the LSS spectrum after the source elimination described in
subsection \ref{subsec:elimination}, in which the reduced fraction of 85.3\%
for the sky coverage is corrected;
(e) the simulated component of the stray light which comes from
the outside of the accumulation radius of 20 mm ($\simeq 20'$)
from the respective optical axes;
(f) the summed spectrum of the resolved sources, i.e.\ $\rm (c) - (d)$.
}\label{fig:uchiwake}
\end{figure}

We estimated that the NXB had been reproduced by an accuracy of 3\% 
using the H0 + H2 monitor count and corrected for the long-term trend.
Therefore, the systematic uncertainty in the NXB subtraction
affects the CXB intensity by at most 10\%,
even at $\sim 7$ keV where the NXB dominates the CXB
(see figure~\ref{fig:uchiwake}).

\subsection{Source Elimination}
\label{subsec:elimination}
The CXB intensity can be defined by equation (\ref{eq:FS}) 
in subsection ~\ref{subsec:Sim}, even including the contribution
from very bright sources, namely, by setting 
$S\rightarrow\infty$ in equation (\ref{eq:FS}) in subsection \ref{subsec:Sim}.
On the other hand, because actual observations are usually biased to
the faint so-called blank sky, the observed intensity
needs a certain correction.
This correction depends on the observed flux range, $S<S_{\rm 0}$,
where it is assured that only the sources with their fluxes fainter
than $S_{\rm 0}$ exist in a sample field.
Therefore, it is essentially required to eliminate {\em resolved} sources
on the source flux basis.
We employed a certain fixed value of
$S_{\rm 0}\simeq 2.0\times 10^{-13}$ \ergs\ (2--10 keV),
and defined that sources brighter than $S_{\rm 0}$ were {\em resolved}
and that the remained X-ray emission was the {\em unresolved} CXB\@.
This procedure is also important in order to overcome
the selection bias of the AMSS fields,
because they are originally intended to observe specific targets.
Since the AMSS and the LSS catalogs are constructed
on the source significance basis,
some sources with low significances but possible high flux
or extended emission could be missed.
We therefore developed an original method for eliminating sources
by slightly modifying the source detection procedure of the AMSS\@.
Throughout this paper, we call this method as ``source elimination'',
which is described in the following paragraphs.

The resolved sources were eliminated
using the observed data, itself, in each field.
We first constructed a mosaic flat-field image $M^{\rm FLAT}$
in the sky coordinates. This is a probability map that is 
 proportional to the intensity of resolved sources,
after subtracting the NXB and the unresolved CXB,
and correcting for the telescope vignetting and the exposure of each pointing.
This image is useful to set a threshold in terms of the X-ray flux.
The $M^{\rm FLAT}$ was created in the 0.7--7 keV band
since the NXB is dominant in the higher energy band.
In order to eliminate sources at the rim region effectively,
events within a radius of 22 mm ($\simeq 22'$) from the optical axis of
each GIS sensor were accumulated, which is by 10\% larger than
the radius of 20 mm ($\simeq 20'$) when we created an energy spectrum.
A single $M^{\rm FLAT}$ image was created for each sample field, by 
summing up all pointings for both GIS~2 and GIS~3
in the  common sky coordinates.

The NXB image to be subtracted was generated for each pointing
and for each sensor, as described in subsection \ref{subsec:NXB}.
With regard to the unresolved CXB image,
we created a template image for each GIS sensor,
which was commonly used for every field and every pointing.
This template was made from a sunlit-Earth image in 1--3 keV
integrated over two years from 1993 June to 1995 June,
because of its uniformness and extreme brightness.
Since its spectrum is much softer than the CXB,
we corrected the radial brightness profile for the XRT vignetting
utilizing a radial profile of the superposed LSS image in 0.7--7 keV,
in which irregularities due to the discrete sources
are sufficiently smeared out.
The normalization of the template was determined
to give 88\% intensity of the LSS field.
Although this is an empirical factor, it is considered to be
close to the unresolved fraction of the X-ray emission in the LSS field.
Details are described by \citet{Ishisaki1996}.  

After subtracting these two images,
the observed image was cross-correlated with the PSF of the XRT+GIS system.
Since the shape of the PSF of the XRT strongly depends on the source
position on the detector, we calculated it by interpolating
the Cyg X-1 images, which were observed with various offset
angles and azimuth angles within a radius of $17'$ from the optical axis
(\cite{Takahashi1995}; \cite{Ikebe1997}).
The vignetting of the XRT was also corrected here.
This series of processes was made for each sensor (2 sensors)
and each pointing ($N_{\rm P}$ pointings),
and afterwards all $2\times N_{\rm P}$ images were summed up
to build a single mosaic image.
Finally, corrected for the exposure time considering
the overlap of multiple pointings,
the $M^{\rm FLAT}$ image was complete.

For the thus-obtained $M^{\rm FLAT}$ image, 
we settled on a certain threshold level
which corresponded to $\sim 2\times 10^{-13}$ \ergs\ (0.7--7 keV),
and searched for peaks above it.
This flux level was chosen because the sources
were detected at a 5--10~$\sigma$ significance,
depending on the distance from the optical axis,
for a typical exposure of 30--40 ks.
If we assumed a power-law spectrum of $\Gamma\simeq 1.7$,
which is typical for the resolved sources,
the threshold flux did not change very much (less than 10\%)
by altering the energy range from 0.7--7 keV into 2--10 keV\@.
When a peak was detected, a circular region was masked out
from the data. The radius of the mask was determined so that the
remaining surface brightness due to the tail of the PSF would become less
than 10\% of the unresolved CXB\@.
This calculation was conducted using equation (\ref{eq:PSF}),
and the typical mask diameters were $9'$ and $14'$ for fluxes of
$2\times 10^{-13}$ \ergs\ and $1\times 10^{-12}$ \ergs, respectively.
By omitting the masked sky regions,
we created an energy spectrum for the remained area by
accumulating events in each field.
Since the outermost region of the GIS f.o.v.\ is dominated by the NXB,
we collected events within a radius of 20 mm ($\simeq 20'$)
from the optical axis of each sensor.

\begin{figure}[htb] 
\begin{center}
\FigureFile(0.3\textwidth,0.5\textwidth){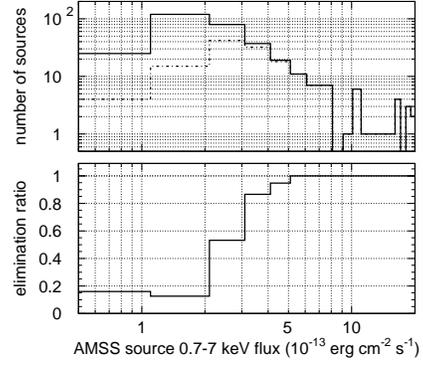}
\end{center}
\caption{Distribution of the 331 AMSS sources in our
analyzed fields as a function of the 0.7--7 keV flux. In the top
panel, the solid line shows all of the AMSS sources in the 91 fields, and
the dotted line indicates the masked-out ones in the source elimination.
The bottom panel shows the elimination ratio for the AMSS sources.
}\label{fig:dist-AMSS}
\end{figure}

\begin{figure}[htb] 
\begin{center}
\FigureFile(0.3\textwidth,0.5\textwidth){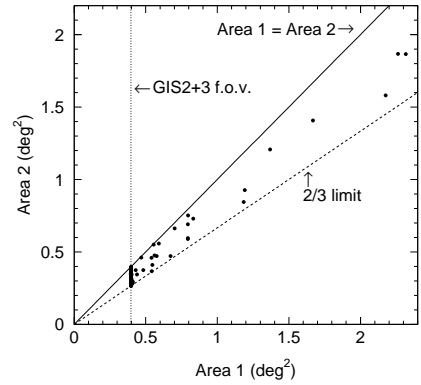}
\end{center}
\caption{Comparison of the sky coverage before
and after the source elimination. The horizontal and vertical axes
correspond to the area before (Area 1) and after (Area 2) the
source elimination, respectively. Three lines which limit the data
scattering are also indicated.
}\label{fig:menseki}
\end{figure}

There are 331 sources listed in the AMSS catalog
in our 88 selected AMSS fields.
As shown in figure~\ref{fig:dist-AMSS},
90\% of those AMSS sources were eliminated in this process
at a flux level of $4\times 10^{-13}$ \ergs\ in 0.7--7 keV\@.
Some fraction of sources fainter than the threshold of
$\sim 2\times 10^{-13}$ \ergs\ (0.7--7 keV) are also eliminated
because they happen to exist neighboring other bright sources.
For each field, the resultant sky area after the source elimination
is compared with the original value in figure~\ref{fig:menseki}.
The sky coverage was thus reduced by about 20\% with this process.

\subsection{Instrumental Response}
\label{subsec:response}

An accurate instrumental (XRT+GIS) response is required
to quantify the CXB spectrum.
However the XRT response depends significantly on the off-axis angle,
making the overall response to the CXB much different
from that to a point source.
The  superposition of point-source responses within the integration region
is still insufficient due to the stray light.
Moreover, the responses are slightly different
among the 91 sample fields,  
since the masking patterns of the source elimination are different
from field to field.

In order to include these conditions in the response,
we performed ray-tracing simulations for all 91 fields using
the SimARF code (\cite{Ishisaki1996}; \cite{Honda1996}; \cite{Shibata2001}),
which calculates the photon detection efficiency
of the XRT+GIS system at each energy,
and generates a so-called Auxiliary Response File (ARF)\@.
In those simulations, we assumed a flat surface brightness
which extends far beyond the GIS f.o.v., to be concrete,
up to $2.^\circ$5 from the optical axis of each sensor.
This assumption is good enough as a first approximation,
at least for the unresolved CXB after the source elimination.
However, it leaves some room for considering
that we can eliminate the resolved sources in the f.o.v.,
while we cannot do so for sources outside the f.o.v.,
from which some fraction of photons appear the stray light.
We deal with this effect in subsection \ref{subsec:CmpStray}.

Though a nuisance, the stray light was reproduced
sufficiently well by a ray-tracing code (\cite{Tsusaka1995}),
which we calibrated against large offset observations
of the Crab nebula up to $\sim 100'$, as shown in figure~\ref{fig:Crab}.
The major origin of the stray light was due to the X-rays to come
through abnormal paths, e.g.\ only one reflection by the primary mirror
or via reflection off the mirror backside,
which were fully taken into account in a ray-tracing simulation.
The stray-light estimation is considered to be accurate to within
$\pm 10$\% of the CXB intensity, with only a mild energy dependence.
We have confirmed that the SimARF generates an identical ARF
for a point source in the f.o.v.\ with that made by ascaarf v2.81
in combination with the standard calibration database files, namely,
{\tt gis2{\rm /}3\_ano\_on\_flf\_180295.fits} (telescope definition file),
{\tt xrt\_ea\_v2\_0.fits} (effective area file),
and {\tt xrt\_psf\_v2\_0.fits} (PSF file).

A so-called {\em ARF filter} was also applied to our SimARF ARF,
which normalized the observed flux of the Crab nebula
with XRT+GIS to the previously reported level
and suppressed small residual structures in the spectral fitting.
With the ARF filter, the Crab spectrum can be expressed by
an absorbed power-law model with Galactic absorption of 
$N_{\rm H}=2.90\times 10^{21}$ cm$^{-2}$, a photon index of $\Gamma=2.09$,
and 2--10 keV flux of $2.16 \times 10^{-8}$ \ergs\ (absorption not corrected).
The ARF filter was applied by default for the ascaarf ARF with
the ``arffil'' parameter; the details are described in \citet{Fukazawa1997}.
As for the energy redistribution matrices,
the released version of {\tt gis2{\rm /}3v4\_0.rmf} were used.
Spectral fits were performed with XSPEC v10.00 \citep{Arnaud1996}.

\subsection{Systematic Errors}
\label{subsec:SysError}

In summary, the constituents of the observed GIS events
in each field can be classified into the following four components:
(1) the unresolved CXB originating in the f.o.v.,
(2) X-rays from outside of the f.o.v.,
(3) the resolved sources to be eliminated, and (4) the NXB\@.
As an example, we plotted each spectrum for the whole LSS field
(including a-LSS) in figure~\ref{fig:uchiwake}.
In this figure, the spectra of $\rm (d)-(e)$, (e), (f), and (b)
correspond to those classifications, respectively.
The LSS field is suitable for this purpose
because of its large sky coverage and the unbiased field selection.
The brightest source in the LSS is AX J131822+3347 with
2--10 keV flux of $1.3\times 10^{-12}$ \ergs.
The count rate of each component in the 0.7--7 (2--10) keV band is
0.055 (0.033), 0.052 (0.022), 0.011 (0.005), and 0.045 (0.047)
count~s$^{-1}$~sensor$^{-1}$, respectively.
Since there is a comparable fraction of the stray light
or the NXB to the unresolved CXB,
accuracy of their estimation, as well as the statistics,
determines the errors of the resultant CXB parameters.

As mentioned in subsection \ref{subsec:NXB},
the reproducibility of the NXB is considered to be 3\%,
which must be carefully dealt with when comparing the
CXB parameters from field to field.
On the other hand, the estimation error of the stray light
mainly acts as if the effective area of the XRT is over- or
under-estimated in common among the sample fields;
hence it does not matter severely in a field-to-field comparison.
As mentioned in subsection \ref{subsec:response},
 because the stray-light estimation is considered to be accurate to 
within $\pm 10$\% of the CXB intensity,
one should only pay attention to comparing our results on
the absolute CXB intensity with those of other satellites.

\section{Spectral Results}
\label{sec:Res}
\subsection{Individual Fields}
\label{sec:Res1}
\begin{figure}[htb] 
\begin{center}
\FigureFile(0.5\textwidth,0.5\textwidth){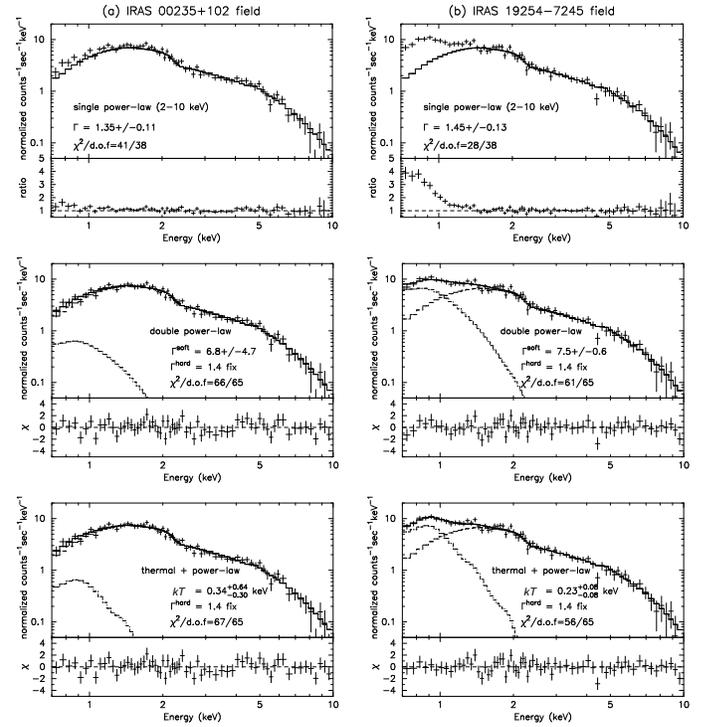}
\end{center}
\caption{Two examples of the energy spectra after the point source
elimination taken in (a) IRAS~00235+102 field and (b) 
IRAS~19254$-$7245 field. The top, middle, and bottom panels show spectral fits
with single power-law, double power-law, and soft thermal plus
power-law models, respectively.  In each panel, the upper diagram shows
the data (cross) fitted with each model (solid lines), and the lower
diagram indicates the goodness of the fit.
}\label{fig:Examples}
\end{figure}

In the first place, all the 91 spectra were fitted with a single
power-law model. The interstellar absorption $N_{\rm H}$ were fixed to
the Galactic values, which range from $0.8 \times 10^{20}$ cm$^{-2}$
to $22.3 \times 10^{20}$ cm$^{-2}$ (\cite{Dickey1990}), and are listed
in table~\ref{tab:obs}. 
When we tried to fit the entire energy band (0.7--10 keV) with the free 
photon index, $\Gamma$, there remained a significant excess below
$\sim 2$ keV in most of the 91 fields.
The reduced $\chi^2$ values showed a progressive improvement
as the low-energy region was gradually cut out from the spectral fit.
Two examples of the spectra are shown in the top
panels of figure~\ref{fig:Examples}. 
Limiting the fitted energy range to be 2--10 keV, an average
power-law photon index of $\Gamma^{\rm hard} = 1.400\pm0.008$
(1~$\sigma$ error)
with a standard deviation of 0.054 was obtained for the 91 sample fields.
This $\Gamma^{\rm hard}$ agrees well with the previous
results from {\it ASCA} (\cite{Gendreau1995}; \cite{Miyaji1998}) and
from non-imaging measurements, such as {\it HEAO~1}\/ A-2
(\cite{Marshall1980}).

The existence of a soft excess has also been noticed in previous 
spectral studies, e.g.\ {\it ROSAT} (\cite{Hasinger1992}),
{\it BBXRT} (\cite{Jahoda1992}), and {\it ASCA} 
(\cite{Gendreau1995}). The present analysis
shows that the excess component is clearly seen in most of the 91
fields in the energy range below 2 keV\@. Such a wide-spread detection
of spectral data and the intensity variation of the soft component,  
as shown in figure~\ref{fig:Examples}, is a newly found feature; 
 the results would supplement the soft X-ray information obtained
with the {\it ROSAT} all-sky survey. 
The variation amplitude of the soft component is
roughly $1\times 10^{-8}$ \ergss.
This is a factor-of-three higher
than the level which can be accounted for by the
change of the Galactic $N_{\rm H}$ between the maximum and the minimum.
To fit the 0.7--10 keV spectra in a consistent way, we need an
additional spectral model for the soft component.  For this, we have
tried (1) power-law and (2) MEKAL models. In both cases, the soft and
the hard components were absorbed by the same Galactic $N_{\rm H}$.

For the double power-law fit, the data were unable to simultaneously
constrain the 2 spectral slopes in some fields. In fact, some of the
fit showed a negative slope for the hard component, which seriously
conflicts with the established value of $\Gamma^{\rm hard}\simeq 1.4$. 
We, therefore, fixed the slope of the hard component at 1.4. The fit
for the 91 spectra then gave an average power-law photon index for the soft 
component, $\Gamma^{\rm soft} = 5.76\pm0.04$ (1~$\sigma$ error)
with a standard deviation of 0.33.
Spectral examples are shown in the middle panels of figure~\ref{fig:Examples},
indicating a significant improvement in the fit for the IRAS
19245$-$7245 field.  The slope of $\Gamma^{\rm soft}\sim 6$, however,
indicates a steep rise in the soft band, and would imply a very strong
emission in the {\it ROSAT} band unless some low-energy cut off is
present. This behavior may have resulted from the power-law modeling of the
soft component, and it seems to complicate the analysis procedure. Our
analysis of the Lockman Hole field, which has been extensively studied
from other satellites, also yielded a rather steep value of
$\Gamma^{\rm soft}=6.0^{+0.7}_{-0.6}$ (90\% confidence level)
compared to $\Gamma \sim 2.1$ obtained from {\it ROSAT} PSPC
(\cite{Hasinger1992}).

As the origin of the soft excess, two possibilities are implied from 
{\it ROSAT} observations (\cite{Kerp1994}). 
One component is probably associated with the Galactic halo,
and the other with the Local Hot Bubble (LHB)\@.
In \citet{Miyaji1998} using both the 
{\it ROSAT} PSPC and the {\it ASCA} GIS+SIS data,
these components were fitted with two thermal models,
and each plasma temperature was derived to 
be $\sim 0.14$ keV and $\sim 0.07$ keV, respectively.
However, it is hard for {\it ASCA} GIS to detect these components separately,
due to its limited sensitivity at the low-energy band. 
We therefore applied one thermal MEKAL model (\cite{Liedahl1995})
for these soft thermal components.
Following the fit with the double power-law model,
we again fixed $\Gamma^{\rm hard}$ at 1.4, $N_{\rm H}$ at the Galactic
values and the metal abundance at 1 solar.
The average temperature $kT$ of the thermal component was obtained to be 
$0.39\pm 0.03$ keV (1~$\sigma$) with a standard deviation of 0.26 keV\@. 
Some examples of the spectral fit are shown in the bottom panels of
figure~\ref{fig:Examples}.

\begin{figure}[htb] 
\begin{center}
\FigureFile(0.3\textwidth,0.5\textwidth){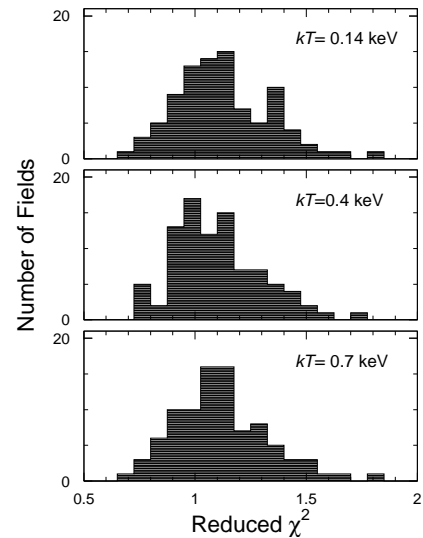}
\end{center}
\caption{Distribution of the reduced $\chi^2$ for spectral fits with three
different models of the soft component.  The plasma temperatures of the soft
thermal component are indicated in each panel.
}\label{fig:kTchi2}
\end{figure}

We then fixed the plasma temperature, $kT$, at several values of between 
0.14 and 0.7 keV\@. The
former value has been indicated by many previous measurements; recent 
examples are \citet{Miyaji1998} mentioned above, and 
{\it ROSAT}/PSPC+rocket CCD experiments (\cite{Mendenhall2001}).
The latter value is implied by {\it BeppoSAX}/LECS (\cite{Parmar1999}), 
although they say that such a high temperature was caused by inadequate 
modeling.
As can be seen in the distribution of reduced $\chi^2$ in 
figure~\ref{fig:kTchi2}, the
temperature range for the soft thermal component is not well
constrained, mainly due to the limitation in the low-energy sensitivity of
GIS\@.  In order to 
examine the field-to-field fluctuation, we used a common $kT$
fixed at the average value, 0.4 keV\@.  
The results of the spectral fits are listed in 
table~\ref{tab:result}. 

\subsection{Integrated Spectrum}

The integrated spectrum for all 91 fields was constructed in order to
look into the fine spectral features. The total integration time
amounted to 4.2~Ms and the spectrum is shown in
figure~\ref{fig:2-10keV}, where the Galactic absorption, $N_{\rm H}$, was 
fixed at $4.0 \times 10^{20}$ cm$^{-2}$, 
which is the average value weighted by the exposure time.
The fitting results for the integrated spectrum are listed in
table~\ref{tab:2-10keV},
as well as those for the whole LSS field (including a-LSS),
because the LSS is a good example of the so-called blank sky.
The average $\Gamma^{\rm hard}$ was obtained to be
$1.411\pm 0.007$ (90\% confidence level).  
The $\chi^2$ value was 175 for 65 degrees of freedom.

The residuals suggest some systematic feature above 8 keV,
characterized by an intensity drop of $\ltsim 30$\%. However, the
systematic error of the NXB intensity is $\sim 3\%$, as mentioned
earlier, and the panels in figure~\ref{fig:2-10keV} show that the
residuals are very much reduced when the NXB level is varied by 
$- 3\%$.
The cutoff feature is, therefore, not significant considering
the error of the NXB level. The night-earth spectrum 
in figure \ref{fig:uchiwake} shows a Cu-K line around 8 keV, which is 
from the gas support grids of GIS\@.
However, this line is clearly subtracted in the integrated spectra.
The edge-like feature around 4.7 keV corresponds to the L-edge
energy of xenon, which is the detector gas of GIS\@. 
In summary, we should say conservatively that the present GIS
data do not indicate any significant deviation of the CXB spectrum from 
the nominal power-law spectrum in the energy range above 2 keV\@.

In order to check the contribution of bright sources in the GIS f.o.v., 
integrated spectra without source elimination were fitted similarly. 
These results are also shown in table~\ref{tab:2-10keV}.
The photon index, $\Gamma^{\rm hard}$, seems not to be changed by applying 
the source elimination, while $F_{\rm X}^{\rm hard}$ is significantly 
affected by $\sim 13$~\% for the sample fields.
This fraction is $\sim 9$\% for the LSS, smaller than the former.
This is because the AMSS fields contains brighter sources,
which are often the target itself, than the LSS\@.
Since the intensities after the source elimination agree well 
with each other,
we can safely state that the source elimination worked well.

\subsection{Correlations between Spectral Parameters}
\label{subsec:COR}

\begin{figure}[htb] 
\begin{center}
\FigureFile(0.5\textwidth,\textwidth){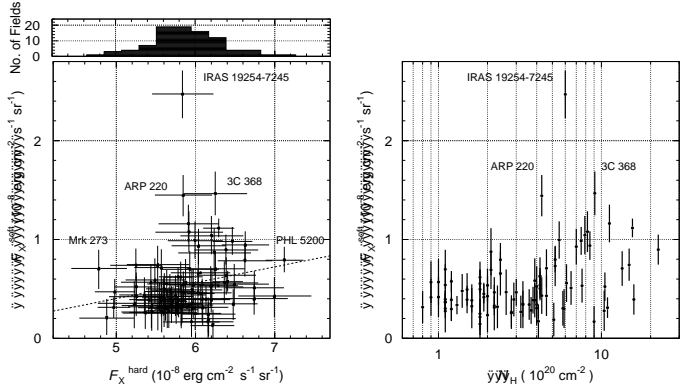}
\end{center}
\caption{The left panel shows correlations between $F_{\rm X}^{\rm hard}$
and $F_{\rm X}^{\rm soft}$, with the distribution histogram of $F_{\rm
X}^{\rm hard}$ plotted in the top panel. The dashed line indicates the
best-fit linear relation. The right panel shows the correlation between
$F_{\rm X}^{\rm soft}$ and $N_{\rm H}$. In both panels, the error bars 
represent the 90\% confidence levels.
}\label{fig:COR}
\end{figure}

\begin{figure}[htb] %Fig. 11
\begin{center}
\FigureFile(0.5\textwidth,0.5\textwidth){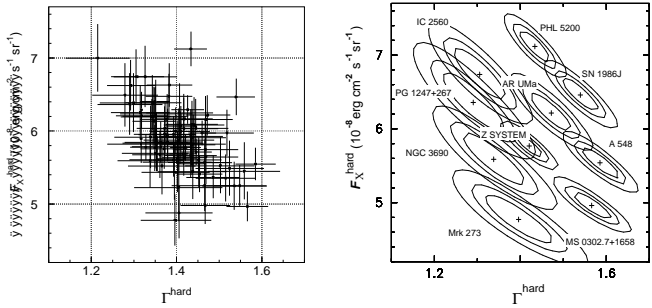}
\end{center}
\caption{The left panel shows correlation between $F_{\rm X}^{\rm hard}$
and $\Gamma^{\rm hard}$, and the right panel shows examples of confidence
contours for several fields.  The error bars represent the 90\% errors for
a single parameter ($\Delta \chi^2=2.70$).
The contours correspond to $\Delta\chi^2$ of 2.3, 4.6, and 9.2.
}\label{fig:COR2}
\end{figure}

The correlations between the spectral parameters obtained in 
subsection \ref{sec:Res1}
 are shown in figures \ref{fig:COR} and \ref{fig:COR2}. 
There seems to be some correlation between
$F_{\rm X}^{\rm soft}$ and $F_{\rm X}^{\rm hard}$ with a large relative
amplitude in the variation of $F_{\rm X}^{\rm soft}$ 
($1~\sigma = 52^{+4}_{-5}~\%$). $F_{\rm X}^{\rm soft}$ does not 
show a negative correlation with $N_{\rm H}$, which
has been observed in very soft X-rays (e.g. \cite{Bowyer1968};  
\cite{Tanaka1977}). On
the other hand, a clear anti-correlation is seen between $\Gamma^{\rm
hard}$ and $F_{\rm X}^{\rm hard}$. The spectral fit for individual
field indicates a systematic correlation between the parameters, as shown
in the right panel of figure~\ref{fig:COR2}, in the sense that
$\Gamma^{\rm hard}$ becomes smaller for a higher $F_{\rm X}^{\rm hard}$. 
However, the total range of the scatter is larger than the statistical
error. We, therefore, conclude there is a systematic tendency between
$F_{\rm X}^{\rm hard}$ and $\Gamma^{\rm hard}$. This correlation
suggests that sources contributing to the fluctuation of the CXB at a
level of $10^{-13}$ \ergs\ have relatively hard spectra.

\section{Large-Scale Distribution}
\label{sec:Large}

The present data are useful for looking at the all-sky distribution of
the CXB with good sensitivity after point-source elimination. We
 examine here the large-scale distribution of both the soft and 
 hard components.

\subsection{Distribution in Our Galaxy}
\label{subsec:sumspec}

\begin{figure}[htb] 
\begin{center}
\FigureFile(0.5\textwidth,0.5\textwidth){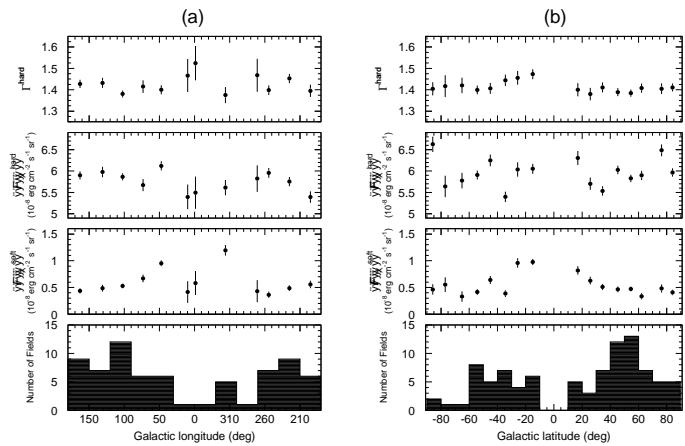}
\end{center}
\caption{Distribution of the spectral parameters in our Galaxy based on
the fits of spatially sorted pulse-height spectra. Left and right
panels are for the galactic longitude ($l$) and latitude ($b$),
respectively. The four rows from top to bottom show $\Gamma^{\rm hard}$,
$F_{\rm X}^{\rm hard}$, $F_{\rm X}^{\rm soft}$, and number of fields,
respectively. The error bars correspond to the 90\% confidence levels.
}\label{fig:BLsort}
\end{figure}

We first consider the spatial distribution of the soft and hard
components in our Galaxy.
The spectrum in each field was sorted
with the Galactic latitude, $b$, and the longitude, $l$,
and summed up in each cell.
The spatial distributions are shown in figure~\ref{fig:BLsort}.
In this plot, the data were binned into $10^\circ$ steps in $b$
and $30^\circ$ steps in $l$, respectively.
In plotting the distribution along $l$, an additional condition of
$|b|<60^\circ$ was also applied. The $N_{\rm H}$ values were averaged
over the respective angular cells. The instrumental responses were also
summed within each cell.

\begin{figure}[htb] 
\begin{center}
\FigureFile(0.3\textwidth,0.5\textwidth){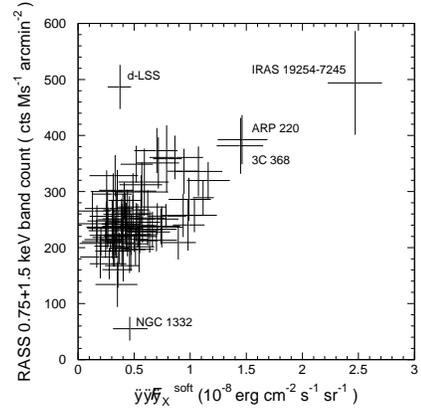}
\end{center}
\caption{Correlation between $F_{\rm X}^{\rm soft}$ and the RASS count
in the 3/4 keV + 1.5 keV band for each sample field.
The RASS count is averaged among the 36$'$ $\times$ 36$'$ 
region centered the {\it ASCA} field.
}\label{fig:RASS-ASCA}
\end{figure}

The soft component shows
a nearly symmetric distribution around a peak at the Galactic Center.
This feature clearly indicates that the soft component is a
Galactic emission, and is likely to extend further into the low-energy
range. To examine this, we compared the soft-component intensity
with the RASS intensity, as shown in figure~\ref{fig:RASS-ASCA}.  The
correlation with the {\it ROSAT}- 3/4 keV + 1.5 keV map
(\cite{Snowden1995}, 1997) is generally good,
suggesting that these enhancements are associated with the Galactic
bulge and the North Polar Spur.
We also looked into some specific fields characterized by strong
soft components. For example, in the IRAS
19254$-$7245 field, the RASS data also indicate a relatively strong flux. 
This supports the above view that the soft component has its origin
within our Galaxy.

In order to compare our results with previous studies, 
we fitted the observed $(l,b)$ profiles of $F_{\rm X}^{\rm hard}$ and
$F_{\rm X}^{\rm soft}$ with a finite radius disk model,
which successfully modeled the 2--60 keV and 2--18 keV distributions
observed with {\it HEAO~1}\/ A-2 (\cite{Iwan1982}) or
{\it Ariel~V}\/ SSI (\cite{Warwick1980}), respectively.
In this model, there is no Galactic X-ray emission outside of
a disk of radius $R_{\rm d}$, and the emission within $R_{\rm d}$ has
an exponential dependence on the vertical distance above the plane
characterized by a scale height parameter, $h$.
The total X-ray intensity, $I_{\rm tot}(l,b)$, is given by
\begin{eqnarray}
I_{\rm tot}(l,b) = I_0 + E\; { h/R_{\rm g} \over \sin |b| } 
\left[ 1 - \exp\left(-{\tan |b| \over h/R_{\rm g}} x \right)\right],
\end{eqnarray}
in which $x = \cos\, l + \sqrt{(R_{\rm d}/R_{\rm g})^2 - \sin^2 l}$,
where $I_0$ is the average isotropic extragalactic emission
assumed in this model, $E$ is the normalization constant for the
Galactic emission, and $R_{\rm g}$ is the distance to the Galactic Center.
When we fit the observed distribution with this model,
additional errors of 5.5\% and 4.0\% times
$F_{\rm X}^{\rm hard}$ were added to $F_{\rm X}^{\rm hard}$ and
$F_{\rm X}^{\rm soft}$, respectively, in order to adjust the reduced
$\chi^2$ unity. This procedure is justified because the observed
extragalactic CXB intensity, itself, would scatter intrinsically
by about this amount, as described in section \ref{sec:logNlogS}.

\begin{figure}[htb] 
\begin{center}
\begin{minipage}{0.24\textwidth}
\leftline{\ \ \small (a) $F_{\rm X}^{\rm soft}$}
\vspace*{-4ex}
\FigureFile(\textwidth,0.25\textwidth){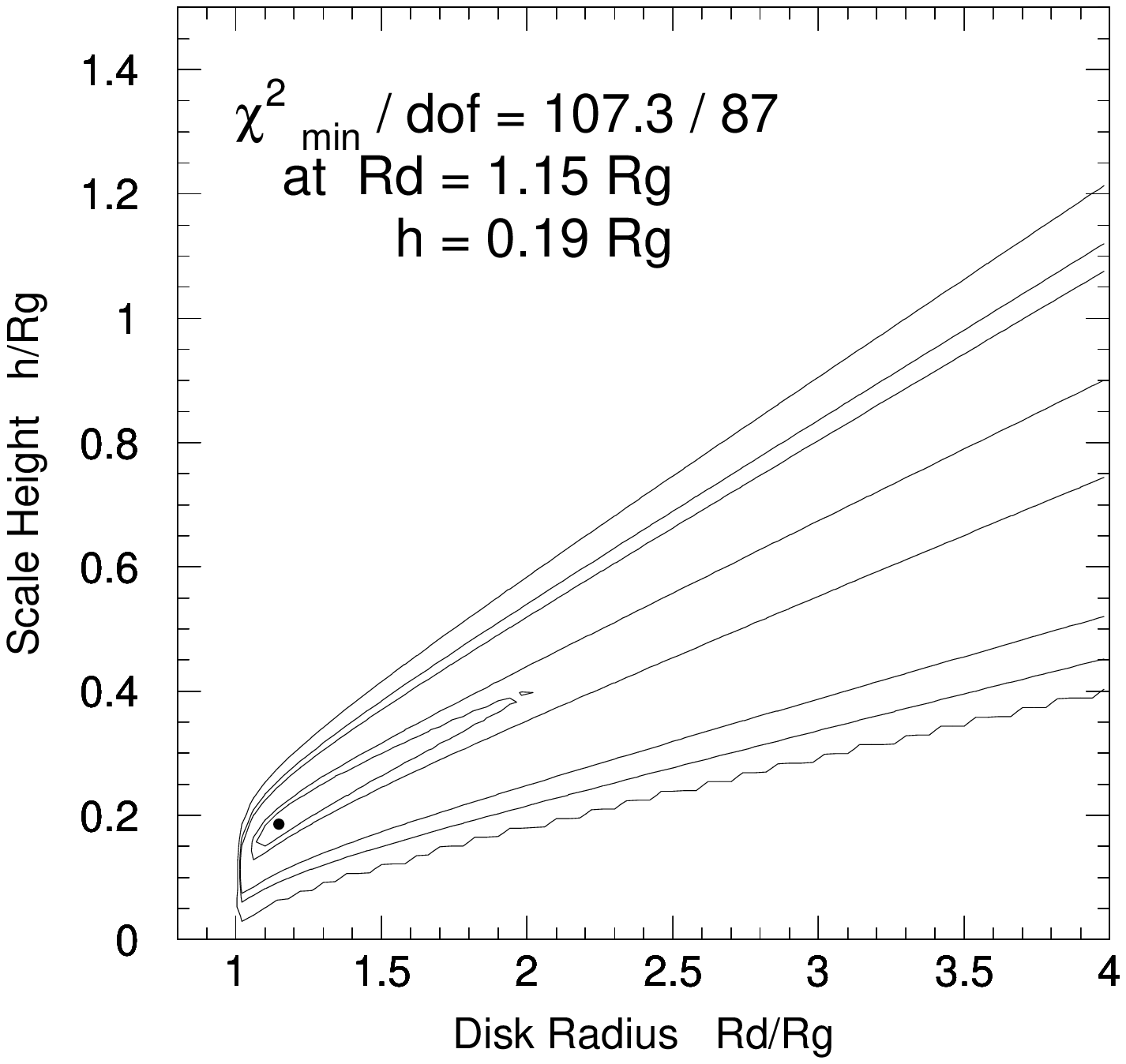}
\end{minipage}
\begin{minipage}{0.24\textwidth}
\leftline{\ \ \small (b) $F_{\rm X}^{\rm hard}$}
\vspace*{-4ex}
\FigureFile(\textwidth,0.25\textwidth){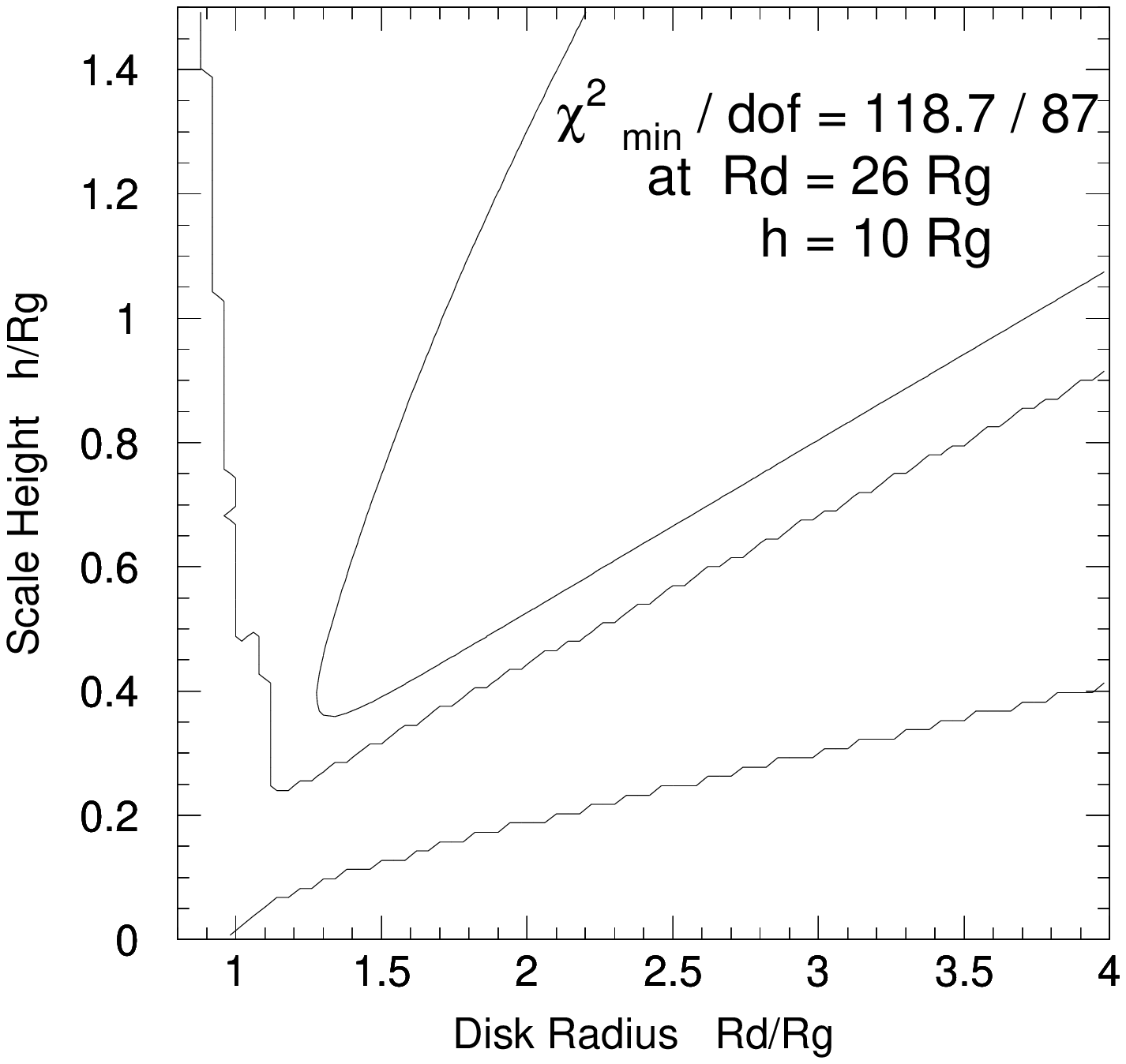}
\end{minipage}
\end{center}
\caption{
Confidence contours between the scale height ($h/R_{\rm g}$) and
disk radius ($R_{\rm d}/R_{\rm g}$) parameters of the finite radius
disk model for the $F_{\rm X}^{\rm soft}$ and $F_{\rm X}^{\rm hard}$.
The contour levels are $\chi^2_{\rm min}+0.5$, +1.0, +4.61 (90\%),
+5.99 (95\%), +9.21 (99\% confidence range), respectively,
from inner to outer. The filled circle in the left panel
represents the $\chi^2_{\rm min}$ position.
For the right panel, the $\chi^2_{\rm min}$ position is out of the panel.
}\label{fig:cont-gal}
\end{figure}

The $F_{\rm X}^{\rm soft}$ distribution was fitted well
by the finite radius disk model, although there is a strong correlation
between the scale height $h$ and the disk radius $R_{\rm d}$,
as shown in figure~\ref{fig:cont-gal}a.
The best-fit parameters are $I_0 = (0.01\pm 0.10)\times 10^{-8}$ \ergss,
$E=(1.9\pm 0.9)\times 10^{-8}$ \ergss, $h=(0.19\pm 0.08)\; R_{\rm g}$,
and $R_{\rm d} = (1.15\pm 0.23)\; R_{\rm g}$ (errors are 1~$\sigma$),
with $\chi^2/{\rm d.o.f.} = 107.3 / 87$.
Since $I_0$ is consistent to be zero and $E\gg I_0$,
we can say that the soft thermal component is almost entirely 
a Galactic emission. The absorption-corrected flux of
the hard power-law component in the 0.5--2 keV energy band
is $2.07\times 10^{-8}$ \ergss\ if we assume the average
photon index of $\Gamma^{\rm hard}=1.412$ over the whole energy range.
Therefore, the contribution of the Galactic component in the 0.5--2 keV band
is calculated to be 49\% at $(l,|b|)=(0^\circ,20^\circ)$,
and 18\% at $|b|=90^\circ$, respectively.

For the hard power-law component, we could hardly constrain the parameters
for the $F_{\rm X}^{\rm hard}$ distribution by fitting it with
the finite radius disk model, as shown in figure~\ref{fig:cont-gal}b.
We therefore fixed the parameters at $R_{\rm d}=2.8 R_{\rm g}$ and
$h = 0.73 R_{\rm g}$, which were the best-fit values for the
{\it HEAO~1}\/ A-2 observation (\cite{Iwan1982}).
These values are inside of the $\chi^2 < \chi^2_{\rm min} + 1$
region in figure~\ref{fig:cont-gal}b.
We then obtained $E = 3.7\pm 2.3$\% of $I_0 = (5.61\pm 0.13)\times 10^{-8}$
\ergss\ (errors are 1~$\sigma$),  which agrees well with the
{\it HEAO~1}\/ A-2 value of $E=3.14$\%.
Hence, the Galactic component in the 2--10 keV band has a low significance.
Using our best-fit values, the Galactic contribution in the 2--10 keV
band was calculated to be 7.1\% at $(l,|b|)=(0^\circ, 20^\circ)$,
and 2.7\% at $|b|=90^\circ$, although the errors are large.
These estimations are again consistent with the very early results from
{\it Ariel~V}\/ SSI (\cite{Warwick1980}), 7.2\% and 2.5\%, respectively.
We can also compute the Galactic contribution to the CXB fluctuation
to be $1~\sigma = 1.2$\%, based on the best-fit model.
This is much smaller than the observed
fluctuation of $1~\sigma = 6.49^{+ 0.56}_{- 0.61}$~\%
(table~\ref{tab:result}).

\subsection{Dipole Fitting}

\begin{figure*}[htb] %Fig. 9.5
\begin{center}
\FigureFile(0.49\textwidth,0.5\textwidth){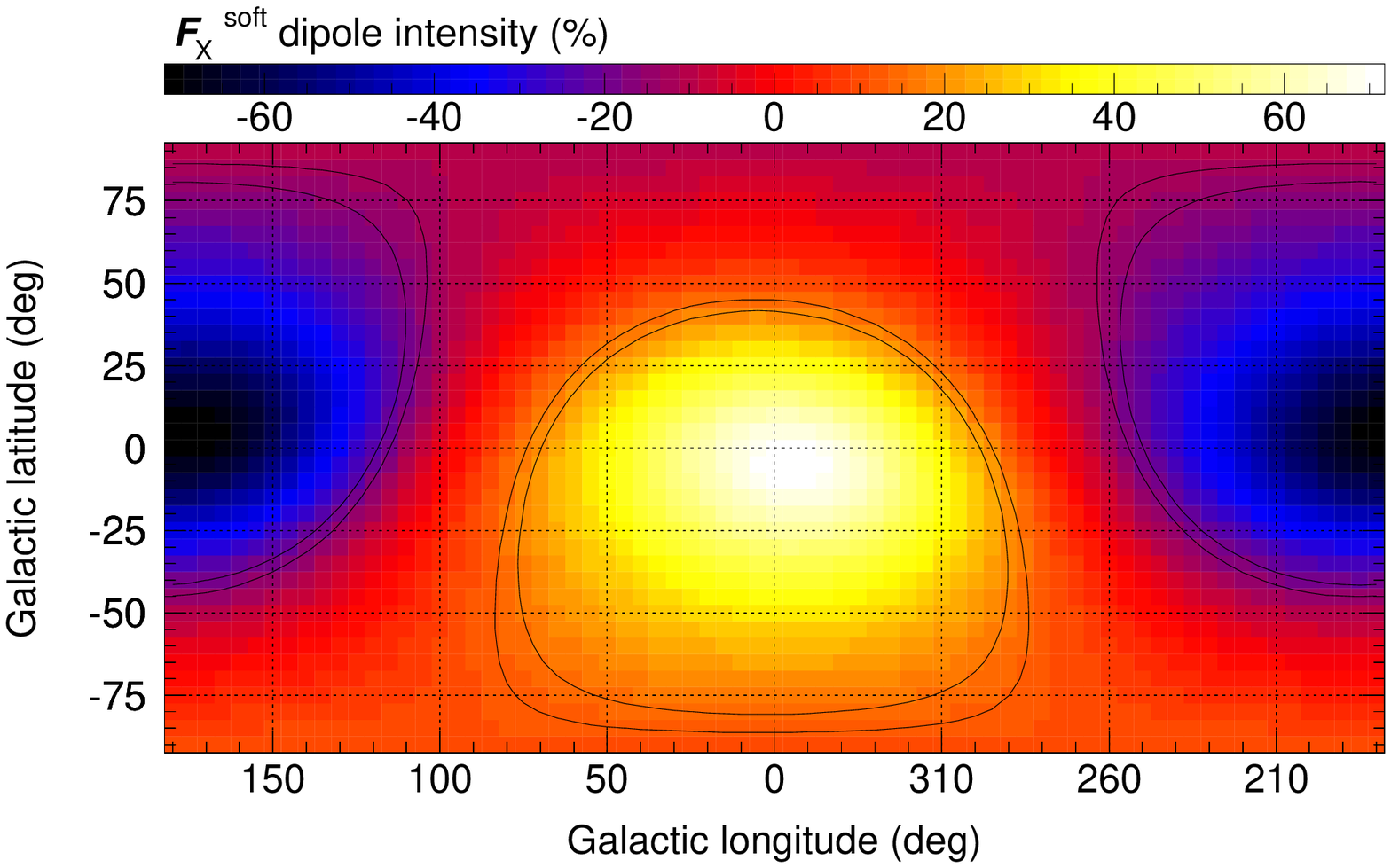}
\FigureFile(0.49\textwidth,0.5\textwidth){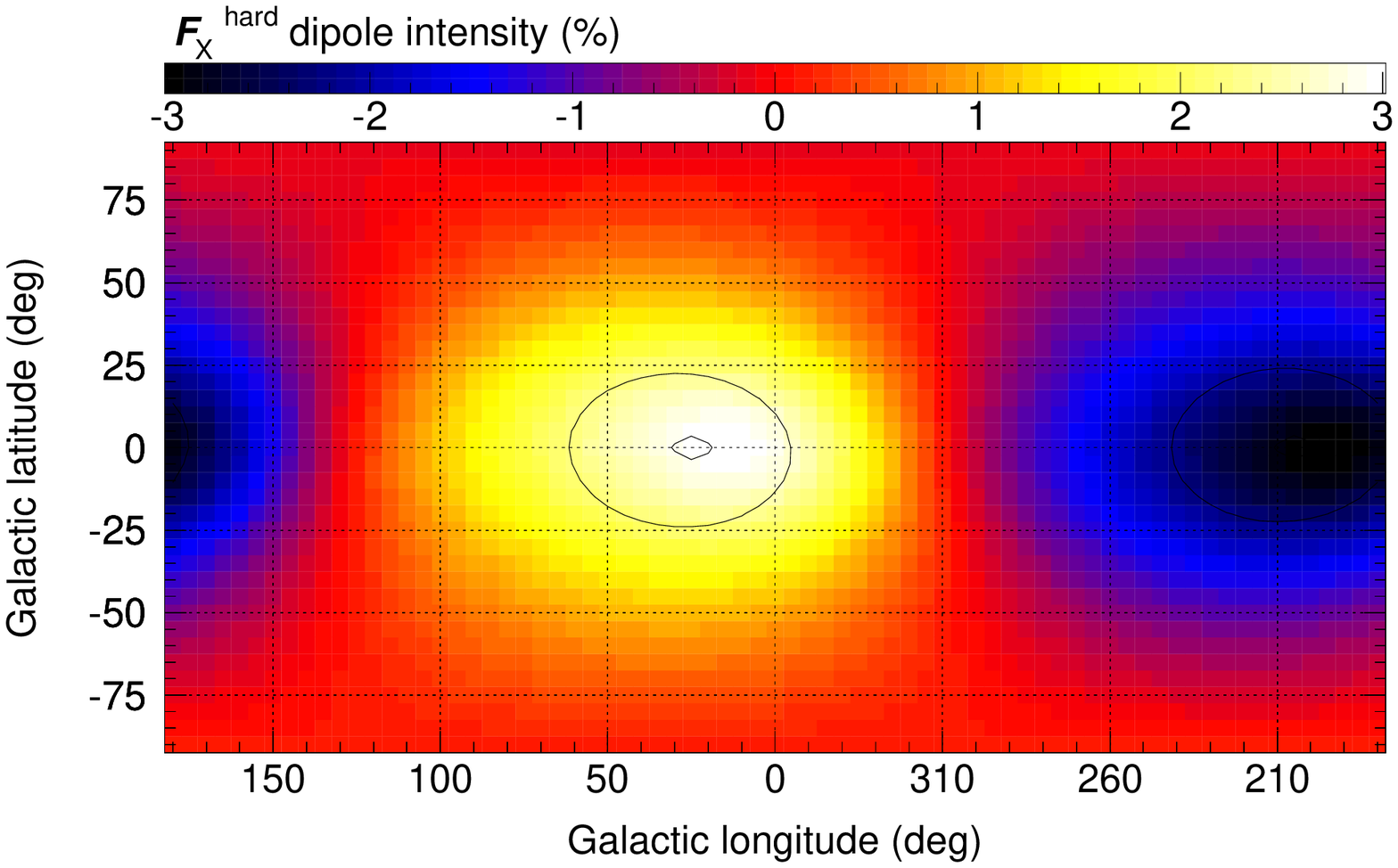}
\end{center}
\caption{Distributions of dipole amplitude for $F_{\rm X}^{\rm soft}$
(left) and $F_{\rm X}^{\rm hard}$ (right) plotted in the Galactic
coordinate. The relative amplitude is indicated by color bars.
The contours correspond to the 90\% and 95\% confidence levels from outer
to inner, where the dipole amplitude has a non-zero value.
}\label{fig:dipole}
\end{figure*}

Aiming at detecting another spatial structure than the Galactic component,
we next tried to fit the intensity distribution with a dipole model,
as a first-order approximation of the spherical harmonics.
The dipole intensity distribution is expressed by
\begin{equation}
\frac{F_{\rm X}(\Theta) - \overline{F_{\rm X}}} 
{\overline{F_{\rm X}}} =I\; \cos\Theta, 
\label{eq:dipole}
\end{equation}
where $\Theta$ is the angle between the center of each field and the
assumed pole direction, $I$ is the amplitude of the dipole,
and $\overline{F_{\rm X}}$ is the average of the observed flux. 
As for the pole direction, we searched for all
directions over the sky with a step size of $5^\circ$.
The hard and soft components were examined separately,
and the errors in the fits were calculated in the same way
as in the previous subsection.
Figure~\ref{fig:dipole} shows the distributions of the best-fit
dipole amplitude in the Galactic coordinate.

As expected from the previous subsection,
the soft component ($F_{\rm X}^{\rm soft}$) 
indicates a strong peak near the Galactic Center
at $(l, b) = (-5^\circ, -5^\circ)$
with a relative amplitude of $72\pm 13\%$ (1~$\sigma$ error).
The hard component ($F_{\rm X}^{\rm hard}$) also suggests some dipole
emission with a maximum amplitude of $3.0\pm 1.6$\% (1~$\sigma$ error)
at $(l, b) = (15^\circ, 0^\circ)$.
There are slight shifts from the Galactic Center
to the positions of the maximum dipole amplitude,
towards the south direction for the $F_{\rm X}^{\rm soft}$
and east for the $F_{\rm X}^{\rm hard}$.
This is presumably due to the influences of the brightest fields,
the IRAS 19254 field at $(l, b) = (322.^\circ 4, -28.^\circ 7)$
for the $F_{\rm X}^{\rm soft}$, and
the PHL 5200 field  at $(59.^\circ 1, -49.^\circ 6)$
for the $F_{\rm X}^{\rm hard}$.
When these fields are removed in the dipole fittings,
the directions of the shift change.

Since the Galactic contribution is too much dominant,
we further tried a dipole fitting after subtracting
the best-fit dipole model with its pole directed to the Galactic Center.
Then, the residuals were consistent with no dipole emission
at the 90\% confidence level in the whole sky,
for both $F_{\rm X}^{\rm soft}$ and $F_{\rm X}^{\rm hard}$.
According to the Cosmic Microwave Background (CMB) dipole
(\cite{Lineweaver1996}), a dipole amplitude of 0.42\%
towards $(l, b) = (264^\circ, 48^\circ)$ is expected
due to the so-called Compton-Getting (CG) effect.
The CG dipole originates in our local motion with respect to the
distant X-ray frame, which is expected to agree with the
direction and velocity determined from the CMB dipole.
There is some evidence that the dipolar emission also exists
in the X-ray band, e.g.\ \citet{Plionis1999} with the 1.5 keV map of
the {\it ROSAT} all-sky survey, \citet{Scharf2000} with the
{\it HEAO~1}\/ A-2 all-sky survey in 2--10 keV\@.
Our dipole amplitudes in this direction after the subtraction
is $-8.1\pm 8.1$\% (1~$\sigma$ error) for the $F_{\rm X}^{\rm soft}$
and $-0.5\pm 1.0$\% (1~$\sigma$ error) for the $F_{\rm X}^{\rm hard}$.
It is almost impossible to detect such a small level of the dipole
($I=0.42$\%) with our data, because the standard deviations of
the observed intensity are much higher (52\% and 6.5\%
for the $F_{\rm X}^{\rm soft}$ and  $F_{\rm X}^{\rm hard}$, respectively)
and the number of sample fields are limited to 91.

\section{CXB Intensity and Fluctuations}
\label{sec:logNlogS}

The present survey consists of 91 fields distributing in the whole sky, 
and provides an excellent database. Regarding the field-to-field fluctuation 
of the hard component ($F_{\rm X}^{\rm hard}$),
we evaluated the contribution from our Galaxy to be 1.2\% 
in the previous section, which is much smaller than the observed
fluctuation of $1~\sigma=6.49^{+ 0.56}_{- 0.61}$~\% 
(table~\ref{tab:result}).
Therefore, the origin of the CXB is considered to be mostly due
to the extra-galactic one,
and its distribution reflects a beam-to-beam
fluctuation of X-ray sources in the observed field.
In this section, we will try
to constrain the \logn\ relation of X-ray sources in the flux
range, $S \ltsim 2\times 10^{-13}$ \ergs\ at 2--10 keV\@.
We compare this fluctuation with simulations, and constrain the 
\logn\ source distribution.

\subsection{Log $N$-- Log $S$ Relation}
\label{subsec:Sim}

In order to examine the
observed fluctuation compared with the expected level from an assumed
\logn\ relation, a detailed simulation 
including all of the instrumental characteristics is essential.
In the {\it ASCA} data, there is a significant flux contribution
($\sim$ 35--45\%; see subsection \ref{subsec:SysError}
from outside of the f.o.v.\ in the observed CXB intensity.
Inside the f.o.v., discrete sources brighter than $S \approx 2\times
10^{-13}$ \ergs\ (2--10 keV) were eliminated,  
whereas we could not exclude brighter sources outside of the f.o.v.\
We only picked up extremely bright sources, which may influence the CXB
intensity by more than 2.5\% based on the
{\it HEAO~1}\/ A-1 catalog and the RASS-BSC,
and discarded the associated fields (subsection \ref{subsec:elimination}).  
We conducted a number of simulations to
effectively constrain the intensity distribution and spectral
properties of the incident sources, including all of 
these systematic effects.

\begin{figure}[htb] 
\begin{center}
\FigureFile(0.5\textwidth,0.5\textwidth){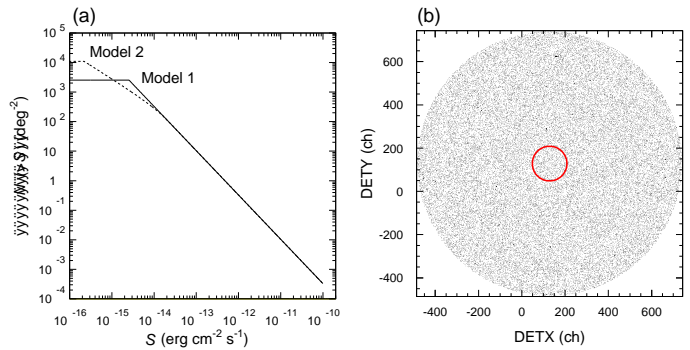}
\end{center}
\caption{Panel (a) shows two assumed \logn\ curves. Model 1 is the
nominal curve ($\gamma = 2.5$) compared with the observed data, and
Model 2 includes the flattening in the low flux range 
($S < 2\times 10^{-14}$ \ergs\ ), as reported
recently.  Panel (b) shows the simulated sample image ($r<2.^\circ 5$) 
following the
\logn\ relation. The central circle represents the GIS field of view,
$r=20$ mm $\simeq 20'$.
}\label{fig:logNlogSsky} 
\end{figure}

We first assumed a certain \logn\ relation which defines
the intensity distribution of X-ray sources in the sky. 
The differential form of the flux ($S$) vs. number ($n$) relation is 
expressed using a normalization $k$ and a slope $\gamma$ as
\begin{equation}
n(S) = -{dN(>S)\over dS} = \left\{ \begin{array}{ll}
0 & (S < S_{\rm min}), \\
kS^{-\gamma} & (S_{\rm min} < S), \\
\end{array}\right.
\end{equation}
where the notation $N(>S)$ means the number density of sources 
brighter than $S$ per steradian. 
The $\gamma$ value equals 2.5 in the 
Euclidean
Universe, which is a good approximation, at least in the range of
$S\gtsim 10^{-13}$ \ergs\ (2--10 keV)\@.  Then, the integrated form is
given as
\begin{equation}\label{eq:lnls}
N(>S) = \left\{ \begin{array}{ll}
{k \over \gamma - 1} S_{\rm min}^{-\gamma+1} & (S < S_{\rm min}), \\
{k \over \gamma - 1} S^{-\gamma+1} & (S_{\rm min} < S). \\
\end{array}\right.
\end{equation}
The CXB intensity in the flux range of $S < S_{\rm 0}$ is calculated as
\begin{equation}\label{eq:FS}
F(S_{\rm 0}) = \int_{S_{\rm min}}^{S_{\rm 0}} n(S)\, S\; dS
 = {k\over \gamma - 2} (S_{\rm min}^{-\gamma+2} - S_{\rm 0}^{-\gamma+2})\,.
\end{equation}
We tentatively adopt the value of $k=k_{\rm 0}=1.58\times 10^{-15}$,
and $\gamma=\gamma_{\rm 0}=2.5$, which gives a reasonable
source density of $N(>S_0)=N_{\rm 0}=1.18\times 10^{4}$ sr$^{-1}$ 
at $S=S_{\rm 0}=2.0\times 10^{-13}$ \ergs\ (2--10 keV)\@.
The $S_{\rm min}$ value is determined to be
$2.52\times 10^{-15}$ \ergs\ (2--10 keV), 
using $F(S_{\rm 0}) = F_{\rm 0} = 5.59\times 10^{-8}$
\ergss\ (2--10 keV) which matches the observed value 
(subsection \ref{subsec:CmpStray}).
In figure~\ref{fig:logNlogSsky}a, the assumed \logn\ relation
with these tentative parameters are shown by the solid line (Model 1).

\subsection{Simulation}
\label{subsec:Sim2}

We next generated a number of skies,
where point sources are randomly spread out up to the off-axis angle
of $2.^\circ 5$, with their source density and flux distribution
following the assumed \logn\ relation.
Figure~\ref{fig:logNlogSsky}b shows an example of the simulated sky.
Among the simulated skies,
we discarded ones with bright sources residing outside of
the GIS f.o.v.\ ($r > 20$ mm $\simeq 20'$),
with the same criterion in selecting the AMSS fields
as indicated in figure~\ref{fig:rass+a1-cut}. 
In order to imitate the source elimination,
we further removed sources inside of the GIS f.o.v.\ 
($r < 20$ mm $\simeq 20'$)
if they were brighter than $S_{\rm 0} = 2.0\times 10^{-13}$
\ergs\ (2--10 keV).
In the actual source elimination, finite regions surrounding these 
eliminated sources were also masked out,
which reduced the averaged detection area by about 20\%.
On the other hand, 33 out of the 91 sample fields consisted of
multi-pointing observations, covering a larger sky area
than that with a single pointing (0.40 deg$^{2}$).
As a result, the sky area covered by one sample field
after source elimination, i.e.\ the mean value of the
``Area 2'' column of table \ref{tab:obs}, is 0.44~deg$^2$ on average,
which is 10\% larger than the integration area for 
a single pointing without any source elimination.

Indeed, these finite sizes of the masking around bright sources
or overlapping multipointing implies non-independent data.
These effects, however, reconcile with a non-uniform exposure,
as long as the source distribution is random,
e.g.\ clustering of sources in a few arcmin scale can be neglected.
We investigated this effect in the simulation
by changing the integration area, $S$, on the detector.
According to equation (\ref{eq:condon}),
a mild $S^{-0.5}$ dependence of the fluctuation width is expected.
Five patterns of the integration area were tested;
namely, we masked out a certain range of azimuth angles, like a pie cutting,
with opening angles of 0$^\circ$, 30$^\circ$, 60$^\circ$, 120$^\circ$, 
and 180$^\circ$.
We found differences in the following simulation results are small and
within statistical errors between 0$^\circ$ (100~\% detector area)
and 60$^\circ$ (83~\%) cases.
Hence, we refer only to the results with a 100\% detector area hereafter.

We lastly generated more than ten millions of photons
for each simulated sky following its intensity map 
in the energy range of 1--12 keV, while assuming that all of the 
sources have a 
common power-law energy spectrum with $\Gamma=1.4$.
The number of photons was chosen so that
it would correspond to a 100 ks exposure with the GIS\@.
We conducted a full instrumental simulation
of the XRT and the GIS for each photon,
which uses basically the same code with the SimARF,
utilizing the ray-tracing and the standard response matrices of
{\tt gis2{\rm /}3v4\_0.rmf}.
Photons detected by the GIS were collected to make an energy spectrum.
Simulations with GIS~2 and GIS~3 were run separately
for a sky, and the resultant two spectra were summed up afterwards.
We performed this instrumental simulation for 369 simulated skies,
and generated a set of spectra of the same number.
Each spectrum was fit by a single power-law model in the 2--10 keV band,
with a common response created by the SimARF in the same way
as described in \S~\ref{subsec:response}, assuming a flat surface brightness.
Note that the NXB was not taken into account in the simulation.
The contribution of the NXB reproducibility is examined separately
in the following subsection, so that we can evaluate the pure Poisson
noise effect in the observed fluctuation.

\subsection{Comparison of Results}
\label{subsec:CmpObsSim}

\begin{figure}[htb] 
\begin{center}
\FigureFile(0.5\textwidth,0.5\textwidth){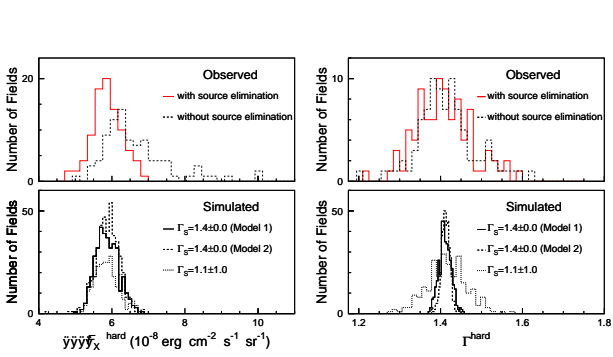}
\end{center}
\caption{Distribution of the 2--10 keV flux $F_{\rm X}^{\rm hard}$
(left panels) and the photon index $\Gamma^{\rm hard}$ (right panels). 
The top row represents the observed distribution, with solid and dotted
histograms showing with and without the source elimination,
respectively.  The bottom panels show simulation results. 
The solid and dotted
lines are for a fixed energy spectrum ($\Gamma_{\rm S} = 1.4\pm 0.0$)
and for the Gaussian distribution, $\Gamma_{\rm S}=1.1\pm 1.0$,
respectively. In both cases, the source elimination was applied.
}\label{fig:SimGFh}
\end{figure}

We thus obtained 369 set of parameters,
$F_{\rm X}^{\rm hard}$ and $\Gamma^{\rm hard}$,
for the assumed \logn\ relation.
The distributions of $F_{\rm X}^{\rm hard}$ and
$\Gamma^{\rm hard}$ for both the observation and the simulation
are shown in figure~\ref{fig:SimGFh}.
In the following comparison, we must take care that
each parameter has an individual error of fitting, i.e.\ statistical error.
We therefore calculated a weighted average, ${\rm Av}(y)$, and its
1~$\sigma$ error, $\delta{\rm Av}(y)$, for a parameter $y$ as
\begin{equation}\label{eq:Av}
{\rm Av}(y) =
\sum_i^{N_{\rm F}} {y_i \over \sigma_i^2}
\left.\right/ 
\sum_i^{N_{\rm F}} {1 \over \sigma_i^2},
\end{equation}
\begin{equation}\label{eq:deltaAv}
\delta{\rm Av}(y) = \sqrt{
{1\over N_{\rm F}-1}
\sum_i^{N_{\rm F}} {[y_i - {\rm Av}(y)]^2 \over \sigma_i^2 }
 \left.\right/ 
\sum_i^{N_{\rm F}} {1 \over \sigma_i^2 }
},
\end{equation}
where $i$, taking between 1 and $N_{\rm F}$, denotes the field ID of
observations or simulations,
and $y_i$ and $\sigma_i$ denote the best-fit value and its 1~$\sigma$ 
error, respectively.
Intrinsic variance ${\rm Sd}^2(y)$ and its 1~$\sigma$ error
$\delta{\rm Sd}^2(y)$, after subtracting the statistical error, 
were computed as
\begin{equation}\label{eq:Sd}
{\rm Sd}^2(y)=
\sum_i^{N_{\rm F}} {[y_i - {\rm Av}(y)]^2 \over \sigma_i^2 }
 \left.\right/ 
\sum_i^{N_{\rm F}} {1 \over \sigma_i^2 }
- N_{\rm F} \left.\right/
\sum_i^{N_{\rm F}} {1 \over \sigma_i^2 },
\end{equation}
\begin{equation}\label{eq:deltaSd}
\delta{\rm Sd}^2(y)=
\sqrt{2\over N_{\rm F}-1}
\sum_i^{N_{\rm F}} {\{y_i - {\rm Av}(y)\}^2 \over \sigma_i^2 }
 \left.\right/ 
\sum_i^{N_{\rm F}} {1 \over \sigma_i^2 }.
\end{equation}
Equations (\ref{eq:Av})--(\ref{eq:deltaSd}) are correct 
as long as both the intrinsic distribution of
the parameter $y$ and the distribution of statistical error of $y_i$
have Gaussian shapes. This assumption seems to be a good approximation
for the parameters after source elimination, as can be seen in
the left-hand panel of figure~\ref{fig:SimGFh}, although this is 
not true for those
without the source elimination where the distribution of 
$F_{\rm X}^{\rm hard}$ have a non-symmetric shape.

With regard to the observed $F_{\rm X}^{\rm hard}$ after the source 
elimination,
${\rm Av}(F_{\rm X}^{\rm hard})$ was calculated to be 
$(5.85\pm0.04)\times10^{-8}$ \ergss\ with a standard deviation of 
$\sqrt{{\rm Sd}^2(F_{\rm X}^{\rm hard})}=0.38^{+0.03}_{-0.04}\times 10^{-8}$
\ergss\ , which corresponds to a $6.49^{+ 0.56}_{- 0.61}$~\% fluctuation.
From the simulation discussed in subsection \ref{subsec:Sim2},
${\rm Av}(F_{\rm X}^{\rm hard})$ was obtained to be 
$(5.77\pm 0.02)\times 10^{-8}$ \ergss\ with fluctuation width 
$5.06^{+0.19}_{-0.20}$~\%. These average values are in good agreement 
between the observation and the simulation,
while the observed fluctuation is larger than that of the simulation.
This discrepancy can be explained by the systematic error,
mainly due to a NXB reproducibility of $\sim 3$\%.
From table~\ref{tab:2-10keV}, the contribution of the systematic error 
to $F_{\rm X}^{\rm hard}$ is estimated to be $\sigma_{\rm S} = 3.2$\%.
Then, the simulation becomes consistent with the observation
at 1~$\sigma$ level, if considering the systematic error.
Therefore, we can say that the Euclidean
distribution of $\gamma = 2.5$ is acceptable with the present data.

As for the observed power-law photon index,  
${\rm Av}(\Gamma^{\rm hard})$ and $\sqrt{{\rm Sd}^2(\Gamma^{\rm hard})}$ 
were obtained to be $1.412\pm 0.007$ and $0.055^{+0.005}_{-0.006}$, 
respectively, 
and ${\rm Av}(\Gamma^{\rm hard}) = 1.414\pm 0.001$ and
$\sqrt{{\rm Sd}^2(\Gamma^{\rm hard})} = 0.009^{+0.001}_{-0.001}$ were 
calculated for the simulated results.
The contribution of the systematic error to $\Gamma^{\rm hard}$
is estimated to be 0.025 from table~\ref{tab:2-10keV}.
We have also examined a contribution due to the large-scale anisotropy
using the dipole analysis, and it is estimated $0.011\pm 0.003$ at most.
Therefore, the observed distribution of $\Gamma^{\rm hard}$ is
wider than that of the simulation,
even though considering the systematic error and the large-scale anisotropy.
This discrepancy is presumably due to the unrealistic assumption that all the
sources have a common power-law index of $\Gamma=1.4$ in the simulation.
We investigate this effect later.

\subsection{CXB Intensity}
\label{subsec:CmpStray}

In order to determine the {\em correct}\/ CXB intensity
in a well-defined flux range,
some compensation for the stray light is required,
because the {\it ASCA} data has a significant flux contribution
from outside of the f.o.v., where we cannot eliminate bright point sources.
In the previous subsection,
we showed that the assumed \logn\ relation
is in good agreement with the observed intensity and fluctuation of 
$F_{\rm X}^{\rm hard}$.
We therefore calculated the {\em correct}\/ CXB intensity using
the assumed \logn\ relation.

Spectral fits for the simulated data indicate that $F_{\rm X}^{\rm hard} 
= (5.77\pm 0.02)\times 10^{-8}$ \ergss. Meanwhile, a simple integration of 
the assumed \logn\ relation, i.e. equation (\ref{eq:FS}), 
gives $F(S_{\rm 0}) = F_{\rm 0} = 5.59\times 10^{-8}$ \ergss\ 
in the flux range $S<S_{\rm 0} = 2.0\times 10^{-13}$ \ergs\ (2--10 keV)\@. 
This is 3.2\% less than the simulation value which includes the contribution
of bright discrete sources residing outside of the f.o.v.\ by chance.
If we integrate this \logn\ curve to $S\rightarrow\infty$, 
$F(\infty)$ becomes $6.29\times 10^{-8}$ \ergss\ (2--10 keV) which is
9.0\% larger than the simulated value. Therefore,
the compensation factor is $- 3.2\%$ in the flux range $S < S_0$ and
$+9.0\%$ for the total CXB flux, respectively. These compensation
factors do not change, even though the \logn\ relation has a break in
the flux range of $S<S_{\rm 0}$, as long as the \logn\ has the same
shape at $S>S_{\rm 0}$, where we think the Euclidean slope of
$\gamma=2.5$ has been almost established. This is because the sources
affecting to the compensation factor have fluxes brighter than $S_0$, 
which by chance lie outside of the f.o.v.

By applying the compensation factor to the observed CXB intensity,
the unresolved CXB flux from sources fainter than $S_{\rm 0}$ were 
reduced to $(5.67\pm 0.04)\times 10^{-8}$ \ergss\ (2--10 keV) when averaged
for the 91 sample fields and corrected for the Galactic absorption.
On the other hand, the total CXB flux integrated over $S\rightarrow\infty$
was calculated to be $(6.38\pm 0.04)\times 10^{-8}$ \ergss.
Note that this value does not include the soft
component, and that the absolute flux determined with {\it ASCA} has 
a systematic error of about 10\%, when compared with the
recent results with {\it Chandra} or {\it XMM-Newton}.
If we convert the latter intensity into a power-law normalization at 1 keV,
assuming a photon index of $\Gamma=1.4$, we obtain $9.66\pm 0.07$ \photonss,
which is entirely consistent with a calculation by \citet{Barcons2000}
of $10.0^{+0.6}_{-0.9}$ \photonss (90\% confidence errors) 
using the {\it ASCA} and {\it BeppoSAX} results.
Though this representation is conventionally used in much CXB literature,
this value is very susceptible to the existence of the soft
component and the power-law index of the hard component.
For example, changing the power-law index, $\Gamma$, by only 0.01 causes a
shift of normalization at 1 keV by as much as 16\%.

\subsection{Constraint on the Log $N$-- Log $S$ Relation}
\label{subsec:ConstlogN-logS}

We then tried to constrain the acceptable range of the
\logn\ relation based on the analytic dependence of
the fluctuation width for a given set of $(k, \gamma)$.
The $S_{\rm min}$ value was determined for each $(k, \gamma)$ pair, 
 using equation (\ref{eq:FS}) with $F(S_{\rm 0}) = 5.67\times 10^{-8}$
\ergss\ (2--10 keV)\@.
According to \citet{Condon1974},
the fluctuation width $\sigma_{\rm F}$ is given analytically as
\begin{equation}\label{eq:condon}
\ifnum1=0
\sigma_{\rm F} \equiv
{\sqrt{{\rm Sd}^2(F_{\rm X}^{\rm hard})} \over {\rm Av}(F_{\rm X}^{\rm hard})}=
{\gamma-2 \over S_{\rm min}^{-\gamma+2} - S_{\rm 0}^{-\gamma+2}}
\sqrt{S_{\rm 0}^{3-\gamma}-S_{\rm min}^{3-\gamma} \over
(3-\gamma)\; k\; \Omega_{\rm eff}},
\else
\sigma_{\rm F} \equiv
{\sqrt{{\rm Sd}^2} \over {\rm Av}}=
{\gamma-2 \over S_{\rm min}^{-\gamma+2} - S_{\rm 0}^{-\gamma+2}}
\sqrt{S_{\rm 0}^{3-\gamma}-S_{\rm min}^{3-\gamma} \over
(3-\gamma)\; k\; \Omega_{\rm eff}},
\fi
\end{equation}
where $\Omega_{\rm eff}$ represents
the effective beam size of the XRT+GIS system.
If we take the values derived from the simulation,
i.e.\ $\sigma_{\rm F} = 5.06$\%, etc.,
$\Omega_{\rm eff}$ is calculated to be 0.516~deg$^2$.
Using this $\Omega_{\rm eff}$ value and given $(k, \gamma)$ pair,
we calculated equation (\ref{eq:condon}) and searched for an acceptable
range in the $(k, \gamma)$ plain.
In the parameter search, we defined $\chi^2(\sigma_{\rm F})$ as
\begin{equation}
\chi^2(\sigma_{\rm F})\equiv
\left\{{
{\rm Sd}^2
 - \left(\sigma_{\rm F} \cdot {\rm Av}\right)^2
 - \left(\sigma_{\rm S} \cdot {\rm Av}\right)^2
\over
\delta{\rm Sd}^2
+ \sqrt{2\over N_{\rm F}-1}
  \left(\sigma_{\rm S} \cdot {\rm Av}\right)^2
}\right\}^2,
\end{equation}
where $\sigma_{\rm S}$ denotes the systematic error of 3.2\%.

\begin{figure}[htb] 
\begin{center}
\FigureFile(0.5\textwidth,0.5\textwidth){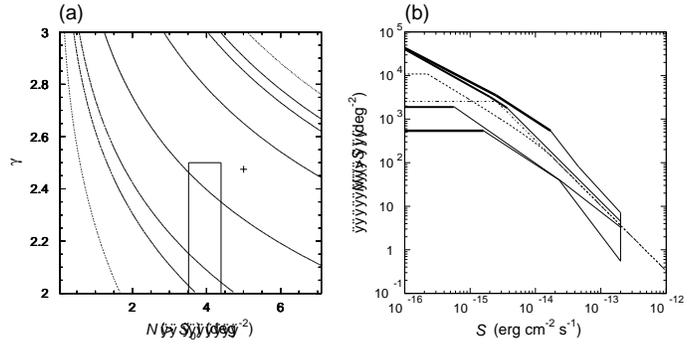}
\end{center}
\caption{In panel (a), the curves indicate the acceptable range in the
$N(>S_0)$ versus $\gamma$ plane. The contours correspond to $\Delta
\chi^2 =$ 1.0 (68.3\%), 4.61 (90\%), 5.99 (95\%),
and 9.21 (99\% confidence range), respectively. The best-fit point is
shown by a cross. The square region shows the constraint derived from
other studies: $\gamma < 2.5$ is given by recent {\it Chandra} and 
{\it XMM-Newton} observations (e.g.\ \cite{Baldi2002, Tozzi2001}), 
and $N(>S_0) = 3.95 \pm 0.43$ deg$^{-2}$ is
estimated from the previous AMSS study (\cite{Ueda1999}). Panel (b)
shows the \logn\ relation matching with the acceptable 
$N(>S_0)$ and $\gamma$ values with two models (dotted lines). 
The outer solid lines correspond to the 90\% confidence limits in panel
(a), and the inner solid lines include the additional constraint by the
square region in (a). In both the outer and the inner lines, 
the thick ones are
boundaries, where the \logn\ corves crosses the 100~\% CXB flux, which we 
determined in section \ref{sec:logNlogS}.
}\label{fig:k-NpN0}
\end{figure}

Since the $\chi^2$ value is determined by 
$\gamma$ and $k$, namely, the degree of freedom equals two,
the 90\% confidence range of $\sigma_{\rm F}$ can be evaluated
as $\chi^2(\sigma_{\rm F}) < 4.61$.
In figure~\ref{fig:k-NpN0}a, we plot the region
where the $(k, \gamma)$ pair gives the 90\% confidence range
of $\sigma_{\rm F}$, as well as that of 68.3\%, 95\% and 99\%.
In this plot, we converted $k$ into $N(>S_0)$,
which represents the $N(>S)$ value for given $(k, \gamma)$ 
at $S=S_{\rm 0}=2\times 10^{-13}$ \ergs.
%% normalized by that with
%% $(k_{\rm 0}, \gamma_{\rm 0}) = (1.58\times 10^{-15},~2.5)$
%%  assumed in the simulation.
The resultant 90\% acceptable region of the \logn\ relation 
is shown in figure~\ref{fig:k-NpN0}b by the outer solid lines.
Roughly speaking, this region is constrained by two factors.
One is a constraint by the fluctuation, which is expressed by
the thin solid lines.
The other is a constraint by the absolute CXB intensity
shown by the thick solid lines, where the integrated source flux
reach 100\% of the absolute CXB intensity,
when the power-law like \logn\ relation is extrapolated
towards the fainter flux range.
In practice, the constraint by the fluctuation is valid
only in the brighter flux range where the Poisson noise, $\sqrt{N}$,
of the source count, $N$, in the f.o.v.\ is comparable to the
fluctuation width, i.e., $N(>S)\ltsim 50$ with our data.
On the other hand, the \logn\ relation must be within the range
of the constraint by the absolute CXB intensity,
as long as the \logn\ curve has a form that is gradually flattening.

Although our obtained region is fairly wide,
we can further constrain the region by rejecting an 
unrealistic set of $(k,~\gamma)$ which contradict with the previous results.
When combined with the LSS and the AMSS results
(\cite{Ueda1999-LSS} and 1999b, respectively),
$N(>S)$ value at $S=S_{\rm 0}$ is determined
in the range of 3.52~deg$^{-2}$ $<$  $N(>S_0)$ $<$ 4.38~deg$^{-2}$. 
%% hence we can constrain $0.9 \le N(>S_0)/N_0 \le 1.3$.
Recent deep observations with {\it Chandra} and {\it XMM-Newton}
show that the flattening of the \logn\ curve comes out 
at a fainter flux around (1--2)~$\times 10^{-14}$ \ergs\ 
(e.g.\ \cite{Baldi2002, Tozzi2001}). Hence, we made another 
condition of $\gamma \le 2.5$. If these two conditions are added,
the acceptable region of the $(k, \gamma)$ pair becomes the area 
surrounded by a rectangle, as shown in figure~\ref{fig:k-NpN0}a, 
and the acceptable \logn\ region becomes the area surrounded by the inner 
solid lines in figure~\ref{fig:k-NpN0}b.
Therefore, the \logn\ relation would reach 100\% of the CXB
at $S<3.7\times 10^{-15}$ \ergs\ if there were no flattening,
and the source density at $S=10^{-16}$ \ergs\ is
in the range of 1900--40000 deg$^{-2}$.

As a confirmation, we tried a simulation while assuming a bending 
\logn\ curve, as 
shown in figures~\ref{fig:logNlogSsky} and \ref{fig:k-NpN0}b (Model 2).
The model was acceptable
with a fluctuation width of $5.48^{+0.19}_{-0.20}$~\%.
We also confirmed the consistency between the analytic formula,
equation (\ref{eq:condon}), by running simulations for the boundary 
values of $k$ and $\gamma$. These results are summarized in table 
\ref{tab:lognlogs}.

\subsection{Spectral Distribution of Sources}
\label{subsec:SpecDist}

As described in subsection \ref{subsec:CmpObsSim},
the simulation could not reproduce the observed deviation of
the CXB spectral index, $\Gamma$, whereas both
the average value and the deviation of $F_{\rm X}^{\rm hard}$
were explained by the simulation fairly well.
This is partly due to the assumption that all of the
sources had a common spectral index of $\Gamma=1.4$ in the simulation.
We therefore investigated the dependence of the apparent CXB spectral index
upon the intrinsic distribution of the source spectra.

In order to investigate this effect,
we introduced a Gaussian distribution of power-law index $\Gamma_{\rm S}$
for those sources which constitute the CXB\@.
We assumed that the distribution of the source spectra is
independent of the source flux.
We chose these assumptions for simplicity,  
although the actual Universe is much more complex.
For instance, hardening of source spectra towards fainter flux range
has already been seen in the AMSS survey (\cite{Ueda1999}),
and the hard sources are usually heavily absorbed sources;
hence, sources with intrinsically flat spectral indices are rare.
However, modeling all of these characteristics is beyond our scope.

The simulation results for several $\Gamma_{\rm S}$ 
are summarized in table~\ref{tab:SimGamma}
along with the observed ones.  The simulation suggests that the intrinsic
deviation of the power-law index, $\Gamma$, is more than 1.0, even when 
considering the deviation due to the systematic error of the NXB which is
estimated to be $\sqrt{{\rm Sd}^2(\Gamma^{\rm hard})}=0.025$  
and the deviation due to the large-scale anisotropy of $0.011\pm0.003$. 
The intrinsic
source spectra must peak at significantly harder index of $\Gamma_{\rm
S}=1.1$ than the observed CXB index of $\Gamma^{\rm hard}\simeq 1.4$ 
if we suppose that their spectral deviation is about 1.0.  This is because
sources with soft spectra emit more photons than do hard sources when
they have the same flux; consequently, the averaged spectrum tends
to be weighted towards the soft source.  On the other hand, we also
found that it was difficult to fit the simulated spectrum with a
single power-law model for a larger deviation of $\Gamma_{\rm S}$,
since the integrated spectrum became a concave shape.
These funny effects are in part caused by the simplistic model,
which is discussed in subsection \ref{subsec:SpecFluct}.

\section{Discussion}
\label{sec:Dis}

The present {\it ASCA} observations have shown interesting features in 
the large-scale distribution of the CXB\@. Below, we will summarize the 
main results and address their implications for each subject.

\subsection{CXB Intensity and Energy Spectrum}

As shown in section \ref{sec:Res}, the integrated energy spectrum of the 
2--10 keV CXB after the source elimination for the total of 4.2~Ms
exposure is described by a power-law model with a photon index
of $\Gamma^{\rm hard}=1.411\pm0.007\pm0.025$ (errors at 90\% statistical 
 and systematic). The slope is consistent with many previous
measurements.  When the nominal NXB is subtracted, the residual from
the power-law model suggests a systematic deviation above 8 keV as if
the observed spectrum has the  cut off as shown in
figure~\ref{fig:2-10keV}\@. However, this feature disappears when the
NXB level is reduced by 3\% within the systematic error.
Therefore, we regard that the cut-off feature above 
8 keV is not significant. Similarly, because the other residual features 
are all
within the NXB uncertainty, we conclude that the energy spectrum
of the CXB is consistent with the nominal power-law model in the
energy range 2--10 keV\@.

The intensity of the power-law component was calculated to be
$8.61\pm0.07$ \photonss\ at 1 keV after eliminating sources with
a flux greater than $S=2\times 10^{-13}$ \ergs. If all the sources
brighter than $2\times 10^{-13}$ \ergs\ were included, the intensity
would become $9.66 \pm0.07$ \photonss\ at 1 keV, using the compensation
factor calculated in subsection \ref{subsec:CmpStray}. This value is 
consistent
with the {\it ASCA} SIS result based on a 250 ks exposure in 4
fields, i.e.\ $9.4\pm 0.4$ \photonss\ (\cite{Gendreau1995}).
Note that these absolute fluxes with {\it ASCA} have systematic errors of
$\sim 10$\% because of the calibration uncertainty.
Nevertheless, our result is considered to be the best estimate
of the CXB intensity so far from two points of view:
a well-calibrated and low-background instrument is used,
and the source elimination ($S_0\sim 2\times 10^{-13}$ \ergs)
and large solid angle ($\Omega_{\rm eff}\sim 50$ deg$^2$)
make the cosmic variance small enough, as expressed
in equation (\ref{eq:condon}).
Compared with results from other missions,
the present one is consistent with the {\it BeppoSAX} LECS value
based on a two thermal plus a power-law model fit;
$10.4^{+1.4}_{-1.1}$ \photonss\ at 1 keV (\cite{Parmar1999}).
However, these values are much smaller than the {\it ROSAT} level of
$13.4\pm0.3$ \photonss\ at 1 keV (\cite{Hasinger1992}), even if we
include the systematic error in the {\it ASCA} data. The relatively
strong {\it ROSAT} flux is caused by the steeper spectrum with
$\Gamma \sim 2.1$, which would be due to contamination of
the soft component or the calibration differences (\cite{Barcons2000}).

\subsection{The Galactic Component}

We found that the $(l,b)$ profiles of both $F_{\rm X}^{\rm soft}$
and $F_{\rm X}^{\rm hard}$ can be fitted with acceptable $\chi^2$ values 
by the finite 
radius disk model, which strongly suggests that a certain fraction of
the X-ray emission has the Galactic origin.
As for the soft thermal component ($F_{\rm X}^{\rm soft}$),
all of the emission is consistent to be Galactic,
with a scale height of $h=1.5\pm 0.6$ kpc and
a disk radius of $R_{\rm d}=9.2\pm 1.8$ kpc
(errors are 1~$\sigma$), if we assume the distance to
the Galactic Center to be $R_{\rm g} = 8$ kpc.
The volume emissivity, $\eta(z_{\rm g})$, is expressed by
\begin{equation}
\eta(z_{\rm g}) = {4\pi E\over h} \exp\left(-{|z_{\rm g}|\over h}\right),
\end{equation}
where $z_{\rm g}$ is the height above the plane.
We can compute the total luminosity of the soft component
by integrating $\eta(z_{\rm g})$ as
\begin{equation}
L_{\rm X} = \int_{-\infty}^\infty \eta(z_{\rm g})\, 
dz_{\rm g} \int_0^{R_{\rm d}} 2\pi r\, dr 
 = 2\, (4\pi E)\, (\pi R_{\rm d}^2).
\end{equation}
Hence, $L_{\rm X}^{\rm soft} = (1.2\pm 0.7)\times 10^{39}$ erg~s$^{-1}$
(0.5--2 keV)\@. \citet{Snowden1997} derived the total luminosity to be 
$\sim 2\times 10^{39}$ erg~s$^{-1}$ from the {\it ROSAT} all-sky survey,
which is consistent with our result.
Note that the absolute value of $F_{\rm X}^{\rm soft}$
is very sensitive to the $kT$ value, which does vary between 0.14 and
0.7 keV, bringing in a systematic error by about an order in
$L_{\rm X}^{\rm soft}$.
Assuming the MEKAL model of the temperature $kT=0.4$ keV
and the metal abundance at one solar, the electron density
in the Galactic plane is calculated as
$n_{\rm e} = (1.5\pm 0.5)\times 10^{-3}$ cm$^{-3}$.

In the same way,
the Galactic component in the hard band ($F_{\rm X}^{\rm hard}$)
is calculated as $L_{\rm X}^{\rm hard}\sim 8\times 10^{38}$ erg~s$^{-1}$
(2--10 keV), although its detection is marginal.
The $F_{\rm X}^{\rm hard}$ also showed a weak dipole feature with
an amplitude of $3.0\pm 1.6$\% and the peak position 
near the Galactic Center, even though the pointed fields
all lie above $|b|=10^\circ$.
This result remained essentially the same
when we simply took the count-rate data above 3 keV
instead of the $F_{\rm X}^{\rm hard}$ obtained from the
spectral fit. The soft component can be approximated by a thermal
model with $kT \sim 0.4$ keV, and even if we take the highest value of 
$F_{\rm X}^{\rm soft}$ detected in the IRAS~19254$-$7245 field,
the flux contribution above 3 keV is only $\sim 0.02$~\%.
This suggests that the distribution of the 2--10 keV X-rays
in our Galaxy really has a high scale-height component.
Essentially the same results were previously reported by 
\citet{Warwick1980} and \citet{Iwan1982}
with the {\it Ariel~V} and {\it HEAO~1}\/ A-2 all-sky surveys, respectively.

\citet{Kokubun2001} recently examined the
spatial distribution of the so-called Galactic bulge and the ridge 
X-ray emission, 
which has a broad enhancement around the Galactic Center and plane,
and showed that the spectrum can be described by a mixture of
two thermal components ($\sim 0.6$ keV and $\sim 3$ keV, respectively)
and a non-thermal power-law component ($\Gamma \sim 1.8$).
Based on an analysis of the {\it ASCA} and {\it RXTE} data,
he suggested the typical scale-height/length of the bulge emission to be
$b\sim 2^\circ$ and $l\sim 7^\circ$.
A comparison with the present result implies that this
emission is extending further away ($|b| > 10^\circ$) from the
Galactic Center with low surface brightness.
\citet{Ebisawa2001} conducted a deep X-ray survey of a region in the Galactic
plane with {\it Chandra} ACIS-I, and found that the hard X-ray emission 
from the Galactic ridge is truly diffuse, not resolved into
discrete sources.
However, there are difficulties in confining plasmas with such
 a energy density and temperature in the Galactic disk.
The far extended hard X-ray emission in our data may indicate
an escape of the high energy plasmas.
The spatial distribution determined by the {\it ROSAT} all-sky survey
gives $h\sim 0.95$ kpc\footnote{
In \citet{Snowden1997}, the scale height parameter is used
for the gas density, hence we have multiplied their scale height by 0.5.}
and $R_{\rm d}\sim 5.6$ kpc, which is slightly smaller than our results
for the $F_{\rm X}^{\rm soft}$.
This is possibly because the {\it ROSAT} PSPC is sensitive in
 a softer band than the {\it ASCA} GIS\@.

\subsection{Log $N$-- Log $S$  Relation}

\begin{figure*}[htb] %Fig. 16
\begin{center}
\FigureFile(\textwidth,\textwidth){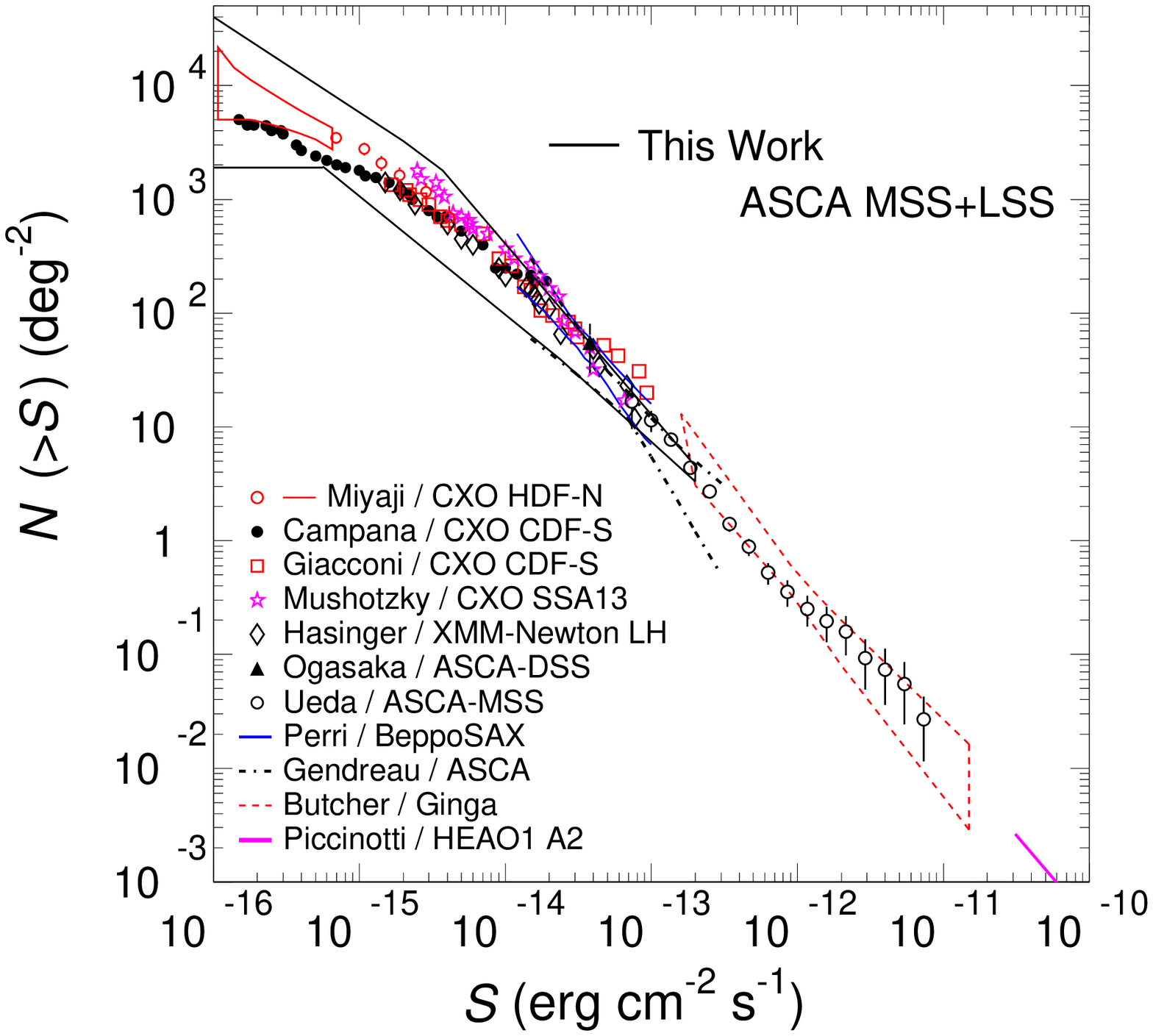}
\end{center}
\caption{\logn\ relation in the energy range 2--10 keV constrained
from the present study (thick solid line), compared with a number of
previous results. The references compiled in this plot are
\citet{Miyaji2002}, 
\citet{Campana2001}, \citet{Giacconi2001}, \citet{Mushotzky2000},
\citet{Hasinger2001}, \citet{Ogasaka1998}, \citet{Ueda1999}, 
\citet{Perri2000}, \citet{Gendreau1998}, \citet{Butcher1997}, and
\citet{Piccinotti1982}.
 }\label{fig:ThelogNlogS}. 
\end{figure*}

Based on the observed fluctuation and the absolute intensity
of the 2--10 keV CXB flux compared with the simulation results,
the acceptable \logn\ relation was constrained as shown in
figure~\ref{fig:ThelogNlogS}\@.  In the flux range above
$\sim 10^{-14}$ \ergs, the relation follows the uniform
Euclidean relation of $dN(>S)/dS \propto S^{-2.5}$, as already
indicated from the previous observations, and lies on a fainter
extension from the previous results from $Ginga$ LAC
(\cite{Butcher1997}) and {\it HEAO~1}\/ A-2 (\cite{Piccinotti1982}).  As
figure~\ref{fig:ThelogNlogS} clearly shows, the present 90\%
confidence range of the \logn\ curve is also consistent with the
fainter-part results obtained from the recent {\it Chandra} and {\it
XMM-Newton} deep-survey observations. We cannot say where the turn
over exactly occurs, but it is close to the $10^{-15}$--$10^{-14}$ 
\ergs\ level from the derived envelope of figure~\ref{fig:ThelogNlogS}.
\citet{Miyaji2002} recently analyzed {\it Chandra} data for the
Hubble Deep Field North with an area of 35.7 arcmin$^2$ by the
fluctuation method.  They derived a loose upper limit in the number
density at the flux level of $2\times10 ^{-16}$
\ergs\ in the 2--10 keV band to be less than 10000 per square degree.
As figure~\ref{fig:ThelogNlogS} shows, the present {\it ASCA} result
in the same energy band gives a consistent upper limit of $\sim 22000$
per square degree, in the sense that an extrapolation of the
\logn\ curve reaches 100\% of the CXB intensity at this point.

The recent {\it Chandra} results also suggest an $\sim 8\%$ difference in
the source contribution brighter than $S>4.5\times 10^{-16}$
\ergs\ between the CDF-N and the CDF-S fields (\cite{Rosati2002}).
However, the measured intensity,
i.e.\ $(5.58\pm 0.56)\times 10^{-8}$ \ergs\ (CDF-N) and
$(5.15\pm 0.49)\times 10^{-8}$ \ergs\ (CDF-S) in 2--10 keV,
have fairly large errors of $\sim 10$\%, and both fields are
restricted to relatively small areas ($\simeq 0.1$~deg$^{2}$).
As pointed out by \citet{Barcons2000}, the cosmic variance plays
an important role in such small-area observations.
We have shown in section \ref{sec:logNlogS} that $\sim 5$\%
of the intensity deviation can be reproduced even by a single \logn\ relation,
and that the fluctuation width is inversely proportional to
the square root of the observed area,
i.e.\ $\sigma_{\rm F}\propto 1/\sqrt{\Omega_{\rm eff}}$ from equation 
(\ref{eq:condon}).
Considering that the GIS+XRT effective beam-size, $\Omega_{\rm eff}$, 
is about 0.5 deg$^2$, it is not unusual that the intensities of
the CDF-N and the CDF-S differ by $\sim 8\%$.
If we take our average value of the CXB intensity  
integrated up to $S\rightarrow\infty$,
i.e.\ $(6.38\pm 0.04)\times 10^{-8}$ \ergss,
the fraction of CXB resolved into discrete sources can be 
$80.8\pm 7.7\%$ in CDF-S and $87.5\pm 8.8\%$ in CDF-N, respectively,
although these values do not take into account the uncertainty
($\sim 10$\%) of the absolute flux with {\it ASCA}\@.

\subsection{Spectral Fluctuation}
\label{subsec:SpecFluct}

As shown in subsection \ref{subsec:CmpObsSim}, the observed distribution of 
$\Gamma^{\rm hard}$ indicates $1~\sigma = 0.055^{+0.005}_{-0.006}$.
If we try to explain this in terms of the Gaussian distribution of
the spectral index of discrete sources,
we need to set the intrinsic fluctuation width to be
larger than 1.0 ($1~\sigma$), and a significant fraction of
the sources must have a harder spectrum than the CXB with an average of
$\Gamma_{\rm S}\sim 1.1$.
This suggests a possibility that faint sources do
not distribute along a simple \logn\ relation, 
but may consist of two or more different populations.
However, the present spectral fluctuation may be partly
coupled with the Galactic structure which affects the observed
variation of $\Gamma^{\rm hard}$ by $\sim 0.011\pm 0.003$ at the maximum.  
Also, the systematic error on the index caused by the NXB 
subtraction is 0.025.
The simulated fluctuation of $\Gamma^{\rm hard}$ is $\pm 0.031$ for a
single \logn\ curve with an intrinsic $\Gamma_{\rm S}=1.1\pm 1.0$; 
therefore, the total fluctuation, including the NXB effect, amounts to 0.041
and approaches the observed level. 

We also note that the Gaussian approximation for the distribution of
the spectral index is not adequate. The actual situation may be that
the distribution broadly consists of two components: one corresponds the
population of unabsorbed AGNs with the peak of index distribution around 
$\Gamma_{\rm S}\simeq 1.7$, 
and the other contains heavily absorbed galaxies with their absorption
ranging over a widely different levels. In fact, several models have
been proposed along with this picture, e.g.\ by \citet{Madau1994},
\citet{Comastri1995}, and \citet{Gilli2001}, and they were successful in
explaining the spectral shape of the CXB\@.
However, the distributions of the intrinsic absorption of AGN 
in these models are based on some assumptions and/or limited observations
of nearby AGNs (e.g.\ 45 Seyferts of $z<0.025$ in
\cite{Risaliti1999}).
Because of the observed spectral fluctuation in our result should include
all of the contribution from faint distant AGNs,
it remains to be studied whether these models can also reproduce
the spectral fluctuation with $1~\sigma\simeq 0.05$
when observed with a solid angle of $\Omega_{\rm eff}\simeq 0.5$ deg$^2$.

\section{Conclusion}
\label{sec:Con}
We measured the absolute CXB intensity and its spectrum from 50 square
degrees based on the low background data with the {\it ASCA} GIS
instrument. A total of 91 selected fields were studied after
eliminating discrete sources with a threshold of
$\sim 2\times 10^{-13}$ \ergs\ (2--10 keV).
The energy spectrum in the energy range of 0.7--10 keV
is well described by a two component model comprizing
a soft thermal emission with $kT\simeq 0.4$ keV and
a hard power-law model with photon index
$\Gamma^{\rm hard}=1.412\pm 0.007\pm 0.025$
(1~$\sigma$ statistical and systematic errors).
As for the extragalactic power-law emission, the intensity fluctuation is
$1~\sigma=6.49^{+ 0.56}_{- 0.61}~\%$,
while the deviation due to the reproducibility of the NXB is 3.2\%.
We also observed a fluctuation in the photon index of 
$0.055^{+ 0.005}_{- 0.006}$, with a NXB contribution of 0.025.

We detected an excess emission toward the Galactic Center
in both the soft (0.7--2 keV) and hard (2--10 keV) energy bands,
which could be fitted by a finite radius disk model.
The emission extends well above $|b| = 10^\circ$.
The soft thermal component is consistent to be totally Galactic,
whereas, for the hard power-law component in the 2--10 keV band, 
the contribution form the Galaxy is 3--7\%.
The soft component showed a strong correlation with the RASS map 
in the 3/4 keV and 1.5 keV bands.

To understand the intrinsic properties of the CXB emission, we carried
out a simulation including the stray-light effect. We found the
absolute CXB intensity to be $(6.38\pm 0.04\pm 0.64)\times 10^{-8}$
\ergss\ (1~$\sigma$ statistical and systematic errors) in the 2--10 keV band.
The observed fluctuation can be explained by the Poisson noise
of the source count in the f.o.v.\ ($\Omega_{\rm eff}\simeq$ 0.5 deg$^2$),
even assuming a single \logn\ relation on the whole sky.
Based on the observed fluctuation and the absolute intensity,
an acceptable region of the \logn\ relation was derived,
and the source density at $S=10^{-16}$ \ergs\ is constrained
in the range of 1900--40000 deg$^{-2}$.
The results turned out to be consistent with
 previous measurements including the recent
{\it Chandra} and {\it XMM-Newton} results.
The spectral fluctuation was modeled with a Gaussian
distribution of photon index, and the simulation indicated that the
intrinsic r.m.s.\ width of $\Gamma$ is larger than 1.0 with a peak
around $\Gamma = 1.1$, suggesting a large amount of hard sources and
a large variation in the intrinsic source spectra.

\bigskip

The authors thank all of the {\it ASCA} team members for their
developing hardware and software, spacecraft operations and
instrumental calibrations. We especially appreciate
the ASCA\_ANL and the SimASCA teams for support in building
the data analysis tools.
We are grateful to Dr.~K.~Kikuchi and Mr.~T.~Kagei for help on
making the NXB database.
We also wish to thank Dr.~K.~Hayashida, Dr.~K.~Masai,
and the referee Dr.~X.~Barcons for their useful comments
and suggestions to improve this manuscript.
This work is supported in part by Grant-in-Aid for Scientific
Research (No.\ 12304009) from the Japan Society for the Promotion of
Science.

\clearpage

\begin{longtable}{cp{7em}cccrrccc}
\caption{Field list.}
\label{tab:obs}
%\begin{center}
\hline\hline
\makebox[2em]{\footnotemark[$*$]Field} & ~~~\footnotemark[$\dagger$]Field & Number of & \footnotemark[$\ddagger$]Observation & \footnotemark[$\S$]$N_{\rm H}$ & 
\multicolumn{1}{c}{\footnotemark[$\|$]$l$} & \multicolumn{1}{c}{\footnotemark[$\|$]$b$} & \makebox[0em]{Exposure} & \footnotemark[$\sharp$]Area 1 & \footnotemark[$**$]Area 2 \\ 
ID & ~~~~name & pointing(s) & date & \makebox[0em]{($10^{20}$ cm$^{-2}$)} & (deg) & (deg) & (ks) & (deg$^2$) & (deg$^2$)  \\
\hline
\endhead
\hline
\endfoot
\hline
\multicolumn{10}{l}{} \\
\multicolumn{10}{l}{\hbox to 0pt{\parbox{180mm}{\footnotesize
\par\noindent
\footnotemark[$*$] Field identification number defined in this paper. 
\par\noindent 
\footnotemark[$\dagger$] Corresponding to the target name in the AMSS 
(Ueda et al. 2001) except for the LSS fields (Ueda et al. 1998).
\par\noindent 
\footnotemark[$\ddagger$] For multi-pointing fields, the earliest observation date are listed.
\par\noindent 
\footnotemark[$\S$] Galactic hydrogen column density toward to the observed field estimated from Dickey and Lockman (1990). 
\par\noindent 
\footnotemark[$\|$] Center of f.o.v.\ in the Galactic coordinates.
\par\noindent 
\footnotemark[$\sharp$] Area before the source elimination.
\par\noindent 
\footnotemark[$**$] Area after the source elimination.
}\hss}}
\endlastfoot
-  & LSS   & \makebox[0em][r]{7}6 & 1993/12/26       & 1.1      & 75.35 & 83.21 & \makebox[0em][r]{5}15.4 & 7.19        & 5.98   \\ \hline
1  & b-LSS & \makebox[0em][r]{2}1 & 1994/01/05       & 1.1      & 73.86 & 83.85 & \makebox[0em][r]{1}30.2       & 2.32  & 1.86   \\ 
2  & c-LSS & \makebox[0em][r]{2}0 & 1994/06/17       & 1.1      & 77.74 & 82.61 & \makebox[0em][r]{1}38.7       & 2.26  & 1.87   \\
3  & d-LSS & \makebox[0em][r]{1}8 & 1995/01/07       & 1.1      & 80.38 & 81.40 & \makebox[0em][r]{1}27.4       & 2.18  & 1.58   \\
4  & DRACO & 3  & 1993/06/04       & 4.2        & 102.28& 33.92 & 57.0  & 0.70  & 0.66   \\
5  & JUPITER & 1  & 1993/06/06     & 2.2        & 286.22& 61.22 & 20.4  & 0.40  & 0.32   \\
6  & NEP~FIELD & 8& 1993/06/09     & 4.2        & 96.38 & 29.90 & \makebox[0em][r]{1}18.5       & 0.58  & 0.47   \\
7  & QSF~3 & 4    & 1993/07/11     & 1.6        & 250.83& $-51.99$& 60.2        & 0.43  & 0.37   \\
8  & ARP~220 & 1  & 1993/07/26     & 4.3        & 36.48 & 53.13 & 23.1  & 0.40  & 0.39   \\
9  & 3C~368 & 1   & 1993/09/12     & 9.1        & 37.57 & 15.27 & 19.9  & 0.40  & 0.36   \\
10 & IRAS~F10214+4724 & 3& 1993/11/04 &1.2      & 168.04& 55.07 & 88.7  & 0.56  & 0.55   \\
11 & IRAS~09104 & 1 & 1993/11/12   & 1.0        & 180.98& 43.66 & 34.5  & 0.40  & 0.27   \\
12 & MNVTEST & 2 & 1993/11/29      & 1.8        & 323.86& $-59.29$& 38.2        & 0.47  & 0.46   \\
13 & SN~1986J & 2& 1994/01/21      & 7.5        & 140.37& $-17.53$& 81.3        & 0.41  & 0.28   \\
14 & AO~0235+164 & 5 & 1994/02/04  & 9.0        & 156.79& $-39.22$& 42.4        & 0.41  & 0.29   \\
15 & NGC~1614 & 1& 1994/02/16      & 6.1        & 204.53& $-34.47$& 28.0        & 0.40  & 0.29   \\
16 & NGC~1667 & 1& 1994/03/06      & 5.5        & 204.13& $-30.21$& 31.0        & 0.40  & 0.33   \\
17 & 4C~41.17 & 1& 1994/03/23      & \makebox[0em][r]{1}0.4     & \makebox[0em][r]{1}74.71& 17.39       & 24.2  & 0.40  & 0.28   \\
18 & NGC~3690 & 2& 1994/04/16      & 0.9        & 141.89& 55.38 & 30.5  & 0.54  & 0.46   \\
19 & K416 & 1    & 1994/05/22      & 1.8        & 97.07 & 62.49 & 24.4  & 0.40  & 0.35   \\
20 & NGC~4449 & 1& 1994/05/25      & 1.4        & 137.20& 72.38 & 37.8  & 0.40  & 0.33   \\
21 & PHL~5200 & 2& 1994/06/21      & 5.2        & 59.13 & $-49.59$& 92.7        & 0.54  & 0.37   \\
22 & NGC~4418 & 1& 1994/06/11      & 2.1        & 289.87& 61.31 & 26.8  & 0.40  & 0.34   \\
23 & GSGP~4  & 4 & 1994/06/24      & 1.9        & 260.39& $-88.46$& \makebox[0em][r]{1}10.5     & 1.37  & 1.21   \\
24 & DI~PEG & 1  & 1994/06/25      & 4.0        & 96.11 & $-43.64$& 24.3        & 0.40  & 0.31   \\
25 & PG~1404+226 & 1 &1994/07/13   & 2.1        & 21.30 & 72.42 & 27.6  & 0.40  & 0.28   \\
26 & IRAS~15307+325 & 1 & 1994/07/22 & 2.0      & 51.95 & 54.92 & 30.1  & 0.40  & 0.39   \\
27 & Z~SYSTEM & 5& 1994/08/09      & 2.2        & 209.43& $-65.08$& \makebox[0em][r]{1}09.9     & 1.67  & 1.41   \\
28 & MG~2016+112 & 2 & 1994/10/24  & \makebox[0em][r]{1}5.5       & 53.88       & $-14.04$& 94.4        & 0.79  & 0.69   \\
29 & IRAS~20460+192 & 1 & 1994/10/27 & \makebox[0em][r]{1}1.2   & 64.55 & $-14.86$& 32.5        & 0.40  & 0.31   \\
30 & BLANK~SKY & 1 & 1994/11/10 & \makebox[0em][r]{1}3.4          & 89.02 & 14.51 & 23.6        & 0.40  & 0.35   \\
31 & EPSILON~CMA & 1 & 1994/11/16 &  \makebox[0em][r]{1}4.8       & 239.85& $-11.22$& 34.4      & 0.40  & 0.31   \\
32 & OJ~287 & 3  &  1994/11/18 &3.0             & 206.80& 35.86 & 88.5  & 0.55  & 0.41   \\
33 & RASS~1011+1736 & 1 & 1994/11/30 & 3.2      & 219.56& 52.58 & 31.8  & 0.40  & 0.39   \\
34 & XY~LEO & 2  &  1994/12/02 & 3.3            & 217.77& 49.85 & 19.0  & 0.40  & 0.31   \\
35 & Mrk~231 & 3 & 1994/12/05 &1.3              & 122.03& 60.27 & \makebox[0em][r]{1}17.1       & 0.83  & 0.73   \\
36 & NGC~7320 & 1&  1994/12/07 & 7.9            & 93.18 & $-21.08$& 28.6        & 0.40  & 0.30   \\
37 & CN~LEO & 1  &  1994/12/16 &2.9             & 244.08& 56.23 & 44.7  & 0.40  & 0.30   \\
38 & Mrk~273 & 1 &  1994/12/27 &1.1             & 107.95& 59.61 & 29.8  & 0.40  & 0.30   \\
39 & NGC~5135 & 1&  1995/01/22 &4.6             & 311.87& 32.50 & 35.6  & 0.40  & 0.28   \\
40 & Q1508+5714 & 2& 1995/03/02 & 1.5           & 93.25 & 51.25 & 63.4  & 0.48  & 0.37   \\
41 & GB~930704 & 3& 1995/03/17 &5.2             & 151.34& 23.89 & 25.0  & 1.18  & 0.84   \\
42 & NGC~4125 & 1& 1995/04/05 & 1.8             & 130.21& 51.27 & 24.8  & 0.40  & 0.30   \\
43 & TXFS~1011+144 & 1  & 1995/05/18 & 3.9      & 224.35& 51.22 & 28.1  & 0.40  & 0.37   \\
44 & MS~1054.5$-$0321 & 1 & 1995/05/23 & 3.6    & 256.54& 48.56 & 48.6  & 0.40  & 0.31   \\
45 & CL~2236$-$04 & 1& 1995/05/31 & 4.0         & 63.00 & $-51.22$& 25.0        & 0.40  & 0.35   \\
46 & PG~1247+267 & 1& 1995/06/17 & 0.9          & 273.35& 89.22 & 26.1  & 0.40  & 0.32   \\
47 & NGC~4450 & 1 & 1995/06/20 &2.4             & 273.70& 78.54 & 25.6  & 0.40  & 0.33   \\
48 & QSO~CLUSTER & 3& 1995/07/03 & 1.1          & 40.01 & 79.08 & \makebox[0em][r]{1}14.8       & 1.19  & 0.93   \\
49 & NGC~5084 & 1 & 1995/07/08 & 8.2            & 311.61& 40.51 & 25.1  & 0.40  & 0.32   \\
50 & SA~68 & 2    & 1995/07/09 & 4.2            & 111.10& $-46.22$& 47.2        & 0.59  & 0.56   \\ 
51 & Mrk~348         & 1 & 1995/08/04 & 5.8     & 122.28& $-30.80$& 32.2        & 0.40  & 0.28   \\
52 & NGC~1332        & 1 & 1995/08/05 & 2.2     & 212.03& $-54.30$& 44.4        & 0.40  & 0.36   \\
53 & PKS~0634$-$205    & 1 & 1995/10/09 & \makebox[0em][r]{2}2.3        & 229.89& $-12.29$& 25.9        & 0.40  & 0.32   \\
54 & IRAS~20551$-$425  & 1 & 1995/10/19 & 3.9   & 358.50& $-40.77$& 25.6        & 0.40  & 0.30   \\
55 & HS~1946+7658    &1  & 1995/10/21 & 7.6     & 109.12& 23.52 & 29.7  & 0.40  & 0.40   \\
56 & RGH~12          &1  & 1995/10/29 & 1.5     & 190.96& 47.33 & 34.9  & 0.40  & 0.34   \\
57 & UKMS~1          &1  & 1995/11/12 & 0.8     & 156.69& 49.55 & 41.0  & 0.40  & 0.32   \\
58 & F568$-$06         &2  & 1995/11/16 & 2.0   & 217.50& 59.50 & 38.8  & 0.40  & 0.29   \\
59 & NGC~7217        &1  & 1995/11/19 & \makebox[0em][r]{1}0.5  & 86.40 & $-19.76$& 62.0        & 0.40  & 0.27   \\
60 & NGC~1672        &1  & 1995/11/23 & 2.3     & 268.82& $-38.88$& 27.3        & 0.40  & 0.27   \\
61 & BJS~855         &1  & 1995/11/27 & 4.0     & 250.20& 49.31 & 32.2  & 0.40  & 0.35   \\
62 & PG~1148+549     &1  & 1995/12/07 & 1.2     & 140.39& 60.39 & 27.4  & 0.40  & 0.31   \\
63 & IRAS~00317$-$2142 &1  & 1995/12/11 & 1.6   & 86.77 & $-83.16$& 27.8        & 0.40  & 0.30   \\
64 & NGC~5005        &1  & 1995/12/13 & 1.1     & 101.12& 79.18 & 27.1  & 0.40  & 0.27   \\
65 & NGC~5018        &1  & 1996/01/16 & 7.0     & 310.05& 43.07 & 28.9  & 0.40  & 0.37   \\
66 & 4C~38.41        &4  & 1996/03/04 & 1.0     & 61.22 & 42.31 & 26.9  & 0.44  & 0.34   \\
67 & AR~UMa          &1  & 1996/04/27 & 1.8     & 167.63& 64.89 & 40.0  & 0.40  & 0.36   \\
68 & PG~1114+445     &1  & 1996/05/05 &1.8      & 164.87& 64.43 & 49.2  & 0.40  & 0.30   \\
69 & HE~1104$-$1805    &1  & 1996/05/31 & 4.6   & 270.78& 37.79 & 30.8  & 0.40  & 0.28   \\
70 & RX~J1716.6+670  &2  & 1996/06/09 & 3.7     & 97.63 & 34.05 & 54.7  & 0.40  & 0.31   \\
71 & B2~1308+326     &2  & 1996/06/10 &1.1      & 86.16 & 83.41 & 28.0  & 0.40  & 0.28   \\
72 & Q2345+007       &2  & 1996/06/26 &3.8      & 92.15 & $-58.00$& 54.4        & 0.40  & 0.34   \\
73 & IRAS~00235+102  &1  & 1996/07/07 & 5.1     & 112.92& $-51.58$& 31.6        & 0.40  & 0.35   \\
74 & NGC~612         &1  & 1996/07/12 & 1.8     & 261.19& $-77.05$& 58.8        & 0.40  & 0.35   \\
75 & CL~0024+17      &1  & 1996/07/21 & 4.2     & 114.54& $-45.20$& 34.6        & 0.40  & 0.29   \\
76 & NGC~315         &1  & 1996/08/05 & 5.9     & 124.57& $-32.39$& 26.6        & 0.40  & 0.33   \\
77 & MS~0302.7+1658  &4  & 1996/08/18 & \makebox[0em][r]{1}0.9  & 162.86& $-34.89$& 98.7        & 0.67  & 0.47   \\
78 & V~471~TAURI     &1  & 1996/08/26 &\makebox[0em][r]{1}5.8   & 172.44& $-27.83$& 36.3        & 0.40  & 0.28   \\
79 & RX~J1802+1804   &1  & 1996/09/30 & 8.5     & 44.00 & 18.82 & 50.0  & 0.40  & 0.35   \\
80 & A548            &2  & 1996/10/10 & 1.9     & 229.79& $-25.21$& 62.4        & 0.79  & 0.59   \\
81 & IRAS~19254$-$7245 &1  & 1996/10/16 & 6.0   & 322.40& $-28.70$& 24.8        & 0.40  & 0.33   \\
82 & IRAS~07598+6508 &1  & 1996/10/29 & 4.3     & 151.10& 32.14 & 31.5  & 0.40  & 0.31   \\
83 & HERC-1          &2  & 1996/11/02 & 2.4     & 75.56 & 34.83 & 45.0  & 0.79  & 0.59   \\
84 & IRAS~08572+3915 &1  & 1996/11/03 & 2.6     & 183.33& 41.10 & 22.7  & 0.40  & 0.31   \\
85 & A851            &3  & 1996/11/16 & 1.2     & 171.64& 48.27 & 81.7  & 0.56  & 0.48   \\
86 & NGC~7130        &1  & 1996/11/18 & 2.0     & 9.78  & $-50.32$& 26.0        & 0.40  & 0.35   \\
87 & 10303+7401      &1  & 1996/11/26 & 4.1     & 134.66& 40.06 & 32.2  & 0.40  & 0.36   \\
88 & 4C~06.41        &2  & 1996/12/13 & 2.8     & 241.14& 52.76 & 67.8  & 0.40  & 0.28   \\
89 & M96             &1  & 1996/12/16 & 2.8     & 234.51& 57.15 & 30.8  & 0.40  & 0.39   \\
90 & IC~2560         &1  & 1996/12/19 & 6.5     & 269.48& 19.12 & 27.2  & 0.40  & 0.37   \\
91 & PG~0043+039     &2  & 1996/12/21 & 3.3     & 121.04& $-58.05$& 51.1        & 0.80  & 0.75   \\
\end{longtable}
    
\begin{longtable}{clrccccc}
\caption{\footnotemark[$*$]Results of the spectral fits.}
\label{tab:result}
\hline\hline
Field ID & \multicolumn{1}{c}{Field name} & \multicolumn{1}{c}{~~\footnotemark[$\dagger$]$N_{\rm H}$} 
& $\Gamma^{\rm hard}$ & 
\footnotemark[$\ddagger$]$F_{\rm X}^{\rm hard}$ & \footnotemark[$\S$]$F_{\rm X}^{\rm soft}$ & $\chi^2$ & d.o.f. \\
\hline
\endhead
\hline
\endfoot
\hline
\multicolumn{8}{l}{\hbox to 0pt{\parbox{140mm}{\footnotesize
\par\noindent
\footnotemark[$*$]
 Results from MEKAL plus power-law model fitting 
(see section \ref{sec:Res}). Error represents 90\% confidence level.   \\
\hspace{1eM}The temperature and abundance are fixed at $kT=0.4$ keV and $Z=1.0$~solar, respectively. 
\par\noindent
\footnotemark[$\dagger$]
 Galactic absorption ($10^{20}$ cm$^{-2})$ for both the power-law and the
MEKAL components. Same as table~\ref{tab:obs}.    
\par\noindent
\footnotemark[$\ddagger$]
 Flux of the power-law component ($10^{-8}$ \ergss) in the 2--10 keV band.
\par\noindent
\footnotemark[$\S$]
 Flux of the MEKAL component ($10^{-8}$ \ergss) in the 0.5--2 keV band.
\par\noindent
\footnotemark[$\|$]
 Weighted average ${\rm Av}$ and its 1~$\sigma$ error $\delta{\rm Av}$ calculated with 
equations (\ref{eq:Av}) and (\ref{eq:deltaAv}).
\par\noindent
\footnotemark[$\sharp$]
 Standard deviation $\sqrt{{\rm Sd}^2}$ and its 1~$\sigma$ error, 
 $\sqrt{{\rm Sd}^2 \pm \delta {\rm Sd}^2} - \sqrt{{\rm Sd}^2}$,
 after the subtraction of statistical  error
 calculated with equations (\ref{eq:Sd}) and (\ref{eq:deltaSd}).\\
}\hss}}
\endlastfoot
1  & b-LSS            & 1.1     &  1.416$^{+0.032}_{-0.032}$ & 5.69$^{+0.16}_{-0.16}$ & 0.40$^{+0.09}_{-0.09}$ & 88.0   & 65 \\
2  & c-LSS            & 1.1     &  1.436$^{+0.031}_{-0.032}$ & 5.55$^{+0.15}_{-0.15}$ & 0.37$^{+0.08}_{-0.09}$ & 65.7   & 65 \\
3  & d-LSS            & 1.1     &  1.446$^{+0.036}_{-0.036}$ & 5.71$^{+0.18}_{-0.18}$ & 0.37$^{+0.10}_{-0.11}$ & 65.9   & 65 \\
4  & DRACO            & 4.2     &  1.393$^{+0.048}_{-0.048}$ & 5.60$^{+0.23}_{-0.23}$ & 0.50$^{+0.12}_{-0.12}$ & 82.8   & 65 \\
5  & JUPITER          & 2.2     &  1.503$^{+0.080}_{-0.083}$ & 5.52$^{+0.39}_{-0.38}$ & 0.32$^{+0.22}_{-0.24}$ & 76.5   & 65 \\
6  & NEP~FIELD        & 4.2     &  1.328$^{+0.037}_{-0.037}$ & 5.78$^{+0.18}_{-0.18}$ & 0.62$^{+0.09}_{-0.08}$ & 104.6  & 65 \\
7  & QSF~3            & 1.6     &  1.342$^{+0.045}_{-0.045}$ & 6.20$^{+0.24}_{-0.24}$ & 0.39$^{+0.12}_{-0.13}$ & 66.5   & 65 \\
8  & ARP~220          & 4.3     &  1.352$^{+0.074}_{-0.074}$ & 5.84$^{+0.37}_{-0.36}$ & 1.45$^{+0.20}_{-0.22}$ & 62.4   & 65 \\
9  & 3C~368           & 9.1     &  1.407$^{+0.079}_{-0.078}$ & 6.25$^{+0.40}_{-0.40}$ & 1.46$^{+0.23}_{-0.22}$ & 96.6   & 65 \\
10 & IRAS~F10214+4724 & 1.2     &  1.349$^{+0.035}_{-0.035}$ & 6.41$^{+0.19}_{-0.19}$ & 0.57$^{+0.11}_{-0.10}$ & 98.2   & 65 \\
11 & IRAS~09104       & 1.0     &  1.215$^{+0.075}_{-0.075}$ & 7.00$^{+0.47}_{-0.46}$ & 0.43$^{+0.19}_{-0.21}$ & 48.0   & 65 \\
12 & MNVTEST          & 1.8     &  1.435$^{+0.056}_{-0.055}$ & 5.81$^{+0.28}_{-0.27}$ & 0.62$^{+0.15}_{-0.17}$ & 67.7   & 65 \\
13 & SN~1986J         & 7.5     &  1.540$^{+0.044}_{-0.047}$ & 6.46$^{+0.25}_{-0.25}$ & 0.98$^{+0.15}_{-0.14}$ & 75.6   & 65 \\
14 & AO~0235+164      & 9.0     &  1.379$^{+0.066}_{-0.069}$ & 5.95$^{+0.36}_{-0.36}$ & 0.17$^{+0.16}_{-0.15}$ & 54.9   & 65 \\
15 & NGC~1614         & 6.1     &  1.395$^{+0.083}_{-0.077}$ & 5.43$^{+0.38}_{-0.38}$ & 0.57$^{+0.18}_{-0.20}$ & 91.7   & 65 \\
16 & NGC~1667         & 5.5     &  1.410$^{+0.068}_{-0.069}$ & 5.99$^{+0.35}_{-0.34}$ & 0.99$^{+0.20}_{-0.18}$ & 90.5   & 65 \\
17 & 4C~41.17         & 10.4    &  1.559$^{+0.084}_{-0.085}$ & 5.45$^{+0.40}_{-0.38}$ & 0.27$^{+0.21}_{-0.19}$ & 73.0   & 65 \\
18 & NGC~3690         & 0.9     &  1.338$^{+0.077}_{-0.077}$ & 5.59$^{+0.38}_{-0.37}$ & 0.42$^{+0.18}_{-0.20}$ & 81.7   & 65 \\
19 & K416             & 1.8     &  1.406$^{+0.078}_{-0.079}$ & 4.88$^{+0.35}_{-0.35}$ & 0.20$^{+0.19}_{-0.17}$ & 62.6   & 65 \\
20 & NGC~4449         & 1.4     &  1.361$^{+0.065}_{-0.065}$ & 5.67$^{+0.33}_{-0.32}$ & 0.47$^{+0.18}_{-0.17}$ & 59.1   & 65 \\
21 & PHL~5200         & 5.2     &  1.434$^{+0.037}_{-0.040}$ & 7.12$^{+0.24}_{-0.24}$ & 0.79$^{+0.14}_{-0.12}$ & 74.7   & 65 \\
22 & NGC~4418         & 2.1     &  1.315$^{+0.069}_{-0.069}$ & 6.25$^{+0.37}_{-0.36}$ & 0.70$^{+0.18}_{-0.20}$ & 70.7   & 65 \\
23 & GSGP~4           & 1.9     &  1.410$^{+0.034}_{-0.033}$ & 6.07$^{+0.17}_{-0.17}$ & 0.45$^{+0.09}_{-0.10}$ & 63.5   & 65 \\
24 & DI~PEG           & 4.0     &  1.345$^{+0.077}_{-0.076}$ & 6.19$^{+0.41}_{-0.40}$ & 0.52$^{+0.21}_{-0.20}$ & 80.0   & 65 \\
25 & PG~1404+226      & 2.1     &  1.351$^{+0.086}_{-0.083}$ & 6.24$^{+0.47}_{-0.45}$ & 0.86$^{+0.25}_{-0.23}$ & 73.8   & 65 \\
26 & IRAS~15307+325   & 2.0     &  1.318$^{+0.061}_{-0.060}$ & 6.21$^{+0.33}_{-0.17}$ & 0.53$^{+0.17}_{-0.16}$ & 47.8   & 65 \\
27 & Z~SYSTEM         & 2.2     &  1.421$^{+0.036}_{-0.036}$ & 5.77$^{+0.18}_{-0.18}$ & 0.33$^{+0.09}_{-0.10}$ & 51.9   & 65 \\
28 & MG~2016+112      & 15.5    &  1.426$^{+0.038}_{-0.038}$ & 6.29$^{+0.20}_{-0.20}$ & 1.11$^{+0.10}_{-0.09}$ & 64.2   & 65 \\
29 & IRAS~20460+192   & 11.2    &  1.436$^{+0.080}_{-0.077}$ & 5.91$^{+0.39}_{-0.38}$ & 1.16$^{+0.19}_{-0.19}$ & 67.3   & 65 \\
30 & BLANK~SKY        & 13.4    &  1.433$^{+0.076}_{-0.074}$ & 5.93$^{+0.37}_{-0.36}$ & 0.70$^{+0.18}_{-0.17}$ & 58.2   & 65 \\
31 & EPSILON~CMA      & 14.8    &  1.535$^{+0.077}_{-0.075}$ & 5.25$^{+0.33}_{-0.32}$ & 0.74$^{+0.17}_{-0.16}$ & 86.1   & 65 \\
32 & OJ~287           & 3.0     &  1.467$^{+0.059}_{-0.058}$ & 4.98$^{+0.26}_{-0.25}$ & 0.47$^{+0.13}_{-0.14}$ & 82.5   & 65 \\
33 & RASS~1011+1736   & 3.2     &  1.433$^{+0.061}_{-0.061}$ & 5.66$^{+0.30}_{-0.29}$ & 0.28$^{+0.15}_{-0.17}$ & 74.1   & 65 \\
34 & XY~LEO           & 3.3     &  1.405$^{+0.103}_{-0.105}$ & 5.22$^{+0.48}_{-0.46}$ & 0.35$^{+0.22}_{-0.25}$ & 78.6   & 65 \\
35 & Mrk~231          & 1.3     &  1.378$^{+0.034}_{-0.033}$ & 5.92$^{+0.17}_{-0.17}$ & 0.34$^{+0.09}_{-0.10}$ & 59.3   & 65 \\
36 & NGC~7320         & 7.9     &  1.350$^{+0.078}_{-0.077}$ & 6.20$^{+0.40}_{-0.39}$ & 1.04$^{+0.20}_{-0.19}$ & 79.9   & 65 \\
37 & CN~LEO           & 2.9     &  1.466$^{+0.064}_{-0.064}$ & 5.57$^{+0.32}_{-0.31}$ & 0.39$^{+0.18}_{-0.17}$ & 61.0   & 65 \\
38 & Mrk~273          & 1.1     &  1.396$^{+0.085}_{-0.080}$ & 4.78$^{+0.36}_{-0.35}$ & 0.71$^{+0.19}_{-0.21}$ & 76.1   & 65 \\
39 & NGC~5135         & 4.6     &  1.452$^{+0.073}_{-0.072}$ & 5.57$^{+0.35}_{-0.34}$ & 0.72$^{+0.18}_{-0.20}$ & 74.3   & 65 \\
40 & Q1508+5714       & 1.5     &  1.347$^{+0.051}_{-0.050}$ & 6.13$^{+0.27}_{-0.27}$ & 0.40$^{+0.13}_{-0.14}$ & 57.5   & 65 \\
41 & GB~930704        & 5.2     &  1.366$^{+0.079}_{-0.080}$ & 5.52$^{+0.40}_{-0.39}$ & 0.75$^{+0.19}_{-0.22}$ & 62.9   & 65 \\
42 & NGC~4125         & 1.8     &  1.372$^{+0.074}_{-0.072}$ & 5.95$^{+0.38}_{-0.37}$ & 0.49$^{+0.21}_{-0.19}$ & 48.7   & 65 \\
43 & TXFS~1011+144    & 3.9     &  1.363$^{+0.069}_{-0.068}$ & 5.79$^{+0.35}_{-0.22}$ & 0.39$^{+0.17}_{-0.18}$ & 70.6   & 65 \\
44 & MS~1054.5$-$0321   & 3.6   &  1.382$^{+0.055}_{-0.055}$ & 6.47$^{+0.31}_{-0.30}$ & 0.35$^{+0.15}_{-0.16}$ & 70.2   & 65 \\
45 & CL~2236$-$04       & 4.0   &  1.349$^{+0.078}_{-0.077}$ & 5.70$^{+0.39}_{-0.24}$ & 0.35$^{+0.18}_{-0.20}$ & 82.4   & 65 \\
46 & PG~1247+267      & 0.9     &  1.290$^{+0.077}_{-0.075}$ & 6.36$^{+0.43}_{-0.42}$ & 0.56$^{+0.22}_{-0.21}$ & 74.0   & 65 \\
47 & NGC~4450         & 2.4     &  1.445$^{+0.077}_{-0.075}$ & 6.04$^{+0.40}_{-0.35}$ & 0.66$^{+0.21}_{-0.23}$ & 69.0   & 65 \\
48 & QSO~CLUSTER      & 1.1     &  1.460$^{+0.033}_{-0.032}$ & 5.97$^{+0.17}_{-0.17}$ & 0.31$^{+0.09}_{-0.11}$ & 83.9   & 65 \\
49 & NGC~5084         & 8.2     &  1.339$^{+0.086}_{-0.085}$ & 5.92$^{+0.43}_{-0.42}$ & 1.07$^{+0.21}_{-0.20}$ & 69.1   & 65 \\
50 & SA~68            & 4.2     &  1.379$^{+0.054}_{-0.052}$ & 5.85$^{+0.27}_{-0.26}$ & 0.54$^{+0.13}_{-0.14}$ & 56.6   & 65 \\ 
51 & Mrk~348         & 5.8      &  1.317$^{+0.079}_{-0.079}$ & 5.82$^{+0.40}_{-0.26}$ & 0.31$^{+0.17}_{-0.19}$ & 78.7   & 65 \\
52 & NGC~1332        & 2.2      &  1.403$^{+0.057}_{-0.058}$ & 5.85$^{+0.30}_{-0.29}$ & 0.46$^{+0.16}_{-0.15}$ & 62.0   & 65 \\
53 & PKS~0634$-$205    & 22.3   &  1.299$^{+0.081}_{-0.079}$ & 6.38$^{+0.41}_{-0.40}$ & 0.89$^{+0.16}_{-0.15}$ & 73.4   & 65 \\
54 & IRAS~20551$-$425  & 3.9    &  1.524$^{+0.081}_{-0.080}$ & 5.49$^{+0.38}_{-0.37}$ & 0.59$^{+0.22}_{-0.23}$ & 63.7   & 65 \\
55 & HS~1946+7658    & 7.6      &  1.518$^{+0.063}_{-0.063}$ & 5.97$^{+0.32}_{-0.31}$ & 0.54$^{+0.17}_{-0.18}$ & 48.7   & 65 \\
56 & RGH~12          & 1.5      &  1.355$^{+0.066}_{-0.065}$ & 5.61$^{+0.32}_{-0.16}$ & 0.49$^{+0.17}_{-0.16}$ & 59.0   & 65 \\
57 & UKMS~1          & 0.8      &  1.408$^{+0.055}_{-0.056}$ & 5.83$^{+0.28}_{-0.28}$ & 0.33$^{+0.15}_{-0.17}$ & 64.1   & 65 \\
58 & F568$-$06         & 2.0    &  1.378$^{+0.060}_{-0.064}$ & 6.16$^{+0.35}_{-0.33}$ & 0.24$^{+0.17}_{-0.19}$ & 60.9   & 65 \\
59 & NGC~7217        & 10.5     &  1.548$^{+0.063}_{-0.062}$ & 5.25$^{+0.28}_{-0.27}$ & 0.52$^{+0.15}_{-0.14}$ & 63.1   & 65 \\
60 & NGC~1672        & 2.3      &  1.438$^{+0.079}_{-0.081}$ & 5.44$^{+0.39}_{-0.38}$ & 0.33$^{+0.20}_{-0.23}$ & 76.1   & 65 \\
61 & BJS~855         & 4.0      &  1.436$^{+0.061}_{-0.061}$ & 6.09$^{+0.32}_{-0.32}$ & 0.31$^{+0.16}_{-0.18}$ & 72.9   & 65 \\
62 & PG~1148+549     & 1.2      &  1.441$^{+0.071}_{-0.071}$ & 6.08$^{+0.38}_{-0.37}$ & 0.40$^{+0.20}_{-0.23}$ & 61.1   & 65 \\
63 & IRAS~00317$-$2142 & 1.6    &  1.439$^{+0.076}_{-0.078}$ & 5.82$^{+0.40}_{-0.39}$ & 0.34$^{+0.21}_{-0.24}$ & 84.0   & 65 \\
64 & NGC~5005        & 1.1      &  1.279$^{+0.078}_{-0.077}$ & 6.49$^{+0.44}_{-0.43}$ & 0.55$^{+0.21}_{-0.22}$ & 72.6   & 65 \\
65 & NGC~5018        & 7.0      &  1.373$^{+0.064}_{-0.066}$ & 6.04$^{+0.35}_{-0.34}$ & 0.94$^{+0.17}_{-0.19}$ & 68.9   & 65 \\
66 & 4C~38.41        & 1.0      &  1.405$^{+0.083}_{-0.082}$ & 5.87$^{+0.43}_{-0.42}$ & 0.56$^{+0.24}_{-0.23}$ & 66.5   & 65 \\
67 & AR~UMa          & 1.8      &  1.472$^{+0.054}_{-0.050}$ & 6.22$^{+0.27}_{-0.28}$ & 0.13$^{+0.16}_{-0.13}$ & 64.5   & 65 \\
68 & PG~1114+445     & 1.8      &  1.398$^{+0.059}_{-0.061}$ & 6.00$^{+0.33}_{-0.31}$ & 0.25$^{+0.16}_{-0.17}$ & 78.1   & 65 \\
69 & HE~1104$-$1805    & 4.6    &  1.467$^{+0.077}_{-0.076}$ & 5.75$^{+0.39}_{-0.38}$ & 0.44$^{+0.20}_{-0.21}$ & 71.8   & 65 \\
70 & RX~J1716.6+670  & 3.7      &  1.371$^{+0.055}_{-0.056}$ & 5.75$^{+0.28}_{-0.27}$ & 0.29$^{+0.13}_{-0.14}$ & 60.1   & 65 \\
71 & B2~1308+326     & 1.1      &  1.402$^{+0.081}_{-0.082}$ & 5.57$^{+0.41}_{-0.41}$ & 0.26$^{+0.23}_{-0.20}$ & 110.8  & 65 \\
72 & Q2345+007       & 3.8      &  1.454$^{+0.054}_{-0.053}$ & 5.41$^{+0.25}_{-0.25}$ & 0.39$^{+0.13}_{-0.14}$ & 61.6   & 65 \\
73 & IRAS~00235+102  & 5.1      &  1.417$^{+0.068}_{-0.064}$ & 6.17$^{+0.34}_{-0.35}$ & 0.20$^{+0.16}_{-0.18}$ & 66.7   & 65 \\
74 & NGC~612         & 1.8      &  1.417$^{+0.052}_{-0.051}$ & 5.63$^{+0.25}_{-0.25}$ & 0.55$^{+0.14}_{-0.13}$ & 95.1   & 65 \\
75 & CL~0024+17      & 4.2      &  1.343$^{+0.068}_{-0.063}$ & 6.39$^{+0.38}_{-0.37}$ & 0.64$^{+0.18}_{-0.20}$ & 65.9   & 65 \\
76 & NGC~315         & 5.9      &  1.469$^{+0.076}_{-0.077}$ & 5.64$^{+0.38}_{-0.37}$ & 0.31$^{+0.19}_{-0.21}$ & 91.2   & 65 \\
77 & MS~0302.7+1658  & 10.9     &  1.566$^{+0.049}_{-0.049}$ & 4.97$^{+0.21}_{-0.20}$ & 0.31$^{+0.10}_{-0.11}$ & 93.6   & 65 \\
78 & V~471~TAURI     & 15.8     &  1.326$^{+0.075}_{-0.074}$ & 6.74$^{+0.43}_{-0.41}$ & 0.40$^{+0.14}_{-0.16}$ & 87.8   & 65 \\
79 & RX~J1802+1804   & 8.5      &  1.383$^{+0.054}_{-0.053}$ & 6.63$^{+0.30}_{-0.30}$ & 0.94$^{+0.14}_{-0.15}$ & 60.9   & 65 \\
80 & A548            & 1.9      &  1.585$^{+0.048}_{-0.047}$ & 5.55$^{+0.23}_{-0.22}$ & 0.41$^{+0.15}_{-0.14}$ & 57.5   & 65 \\
81 & IRAS~19254$-$7245 & 6.0    &  1.353$^{+0.081}_{-0.080}$ & 5.83$^{+0.39}_{-0.38}$ & 2.47$^{+0.24}_{-0.25}$ & 61.8   & 65 \\
82 & IRAS~07598+6508 & 4.3      &  1.515$^{+0.070}_{-0.070}$ & 5.35$^{+0.33}_{-0.32}$ & 0.44$^{+0.18}_{-0.19}$ & 68.2   & 65 \\
83 & HERC-1          & 2.4      &  1.292$^{+0.059}_{-0.055}$ & 6.62$^{+0.35}_{-0.34}$ & 0.78$^{+0.18}_{-0.16}$ & 86.7   & 65 \\
84 & IRAS~08572+3915 & 2.6      &  1.417$^{+0.081}_{-0.080}$ & 6.03$^{+0.42}_{-0.41}$ & 0.46$^{+0.24}_{-0.22}$ & 58.6   & 65 \\
85 & A851            & 1.2      &  1.402$^{+0.045}_{-0.045}$ & 5.62$^{+0.23}_{-0.22}$ & 0.30$^{+0.11}_{-0.13}$ & 70.4   & 65 \\
86 & NGC~7130        & 2.0      &  1.466$^{+0.077}_{-0.077}$ & 5.25$^{+0.36}_{-0.35}$ & 0.42$^{+0.21}_{-0.20}$ & 67.2   & 65 \\
87 & 10303+7401      & 4.1      &  1.468$^{+0.064}_{-0.059}$ & 6.16$^{+0.32}_{-0.33}$ & 0.17$^{+0.18}_{-0.17}$ & 77.8   & 65 \\
88 & 4C~06.41        & 2.8      &  1.452$^{+0.054}_{-0.055}$ & 5.60$^{+0.27}_{-0.26}$ & 0.25$^{+0.14}_{-0.14}$ & 74.1   & 65 \\
89 & M96             & 2.8      &  1.486$^{+0.062}_{-0.063}$ & 5.36$^{+0.30}_{-0.29}$ & 0.26$^{+0.18}_{-0.16}$ & 72.9   & 65 \\
90 & IC~2560         & 6.5      &  1.305$^{+0.066}_{-0.059}$ & 6.74$^{+0.37}_{-0.37}$ & 0.51$^{+0.17}_{-0.17}$ & 64.8   & 65 \\
91 & PG~0043+039     & 3.3      &  1.393$^{+0.049}_{-0.049}$ & 5.95$^{+0.25}_{-0.25}$ & 0.36$^{+0.13}_{-0.14}$  & 86.7  & 65 \\ 
\hline
-- & \multicolumn{1}{l}{\footnotemark[$\|$]{average}} & --~ &  1.412$^{+0.007}_{-0.007}$ & 5.85$^{+0.04}_{-0.04}$ & 0.51$^{+0.03}_{-0.03}$      &  --   &  --  \\  
-- & \multicolumn{1}{l}{\footnotemark[$\sharp$]{standard deviation}} & --~ &  0.055$^{+0.005}_{-0.006}$ & 0.38$^{+0.03}_{-0.04}$ & 0.26$^{+0.02}_{-0.02}$       & -- & -- \\
\end{longtable}

\begin{table*}[htb]
\caption{
Fitting results for the integrated spectrum in the 91 sample fields
and the whole LSS field (including a-LSS).
}\label{tab:2-10keV}
\begin{center}
\begin{tabular}{rcccc}\hline \hline
NXB  & $\Gamma$ & $F_{\rm X}^{\rm hard}$ (2--10 keV) & $F_{\rm X}^{\rm soft}$ (0.5--2 keV) & $\chi^2$/d.o.f. \\
(\%) &  & (10 $^{-8}$ \ergss) & (10 $^{-8}$ \ergss) & \\
\hline
\multicolumn{3}{l}{Integrated spectrum with source elimination} & & \\ 
$+ 3 $ & $1.439^{+ 0.007}_{- 0.007}$ (1.4 fix) & $5.72^{+ 0.04}_{- 0.04}$ ($5.90^{+ 0.02}_{- 0.02}$) & 
$0.49^{+ 0.02}_{- 0.02}$ ($0.55^{+ 0.01}_{- 0.02}$) & 196/65 ~(275/66)\\ 
   0   & $1.411^{+ 0.007}_{- 0.007}$ (1.4 fix) & $5.92^{+ 0.04}_{- 0.04}$ ($5.97^{+ 0.02}_{- 0.02}$) & 
$0.53^{+ 0.02}_{- 0.02}$ ($0.55^{+ 0.01}_{- 0.02}$) & 175/65 ~(182/66)  \\
$- 3 $ & $1.386^{+ 0.006}_{- 0.007}$ (1.4 fix) & $6.11^{+ 0.04}_{- 0.04}$ ($6.04^{+ 0.02}_{- 0.02}$) & 
$0.57^{+ 0.02}_{- 0.02}$ ($0.54^{+ 0.02}_{- 0.01}$) & 199/65 ~(211/66)     \\ \hline 
\multicolumn{3}{l}{Integrated spectrum without source elimination} & &  \\ 
$+ 3 $ & $1.431^{+ 0.006}_{- 0.006}$ (1.4 fix) & $6.61^{+ 0.03}_{- 0.03}$ ($6.77^{+ 0.02}_{- 0.02}$) & 
$0.61^{+ 0.02}_{- 0.02}$ ($0.66^{+ 0.01}_{- 0.01}$) & 243/65 ~(322/66)  \\ 
   0   & $1.407^{+ 0.006}_{- 0.005}$ (1.4 fix) & $6.80^{+ 0.03}_{- 0.03}$ ($6.84^{+ 0.02}_{- 0.02}$) & 
$0.65^{+ 0.02}_{- 0.02}$ ($0.66^{+ 0.01}_{- 0.01}$) & 239/65 ~(243/66)  \\
$- 3 $ & $1.385^{+ 0.006}_{- 0.006}$ (1.4 fix) & $6.99^{+ 0.03}_{- 0.03}$ ($6.91^{+ 0.02}_{- 0.02}$) & 
$0.68^{+ 0.02}_{- 0.02}$ ($0.66^{+ 0.01}_{- 0.01}$) & 286/65 ~(306/66)  \\ \hline
\multicolumn{3}{l}{LSS spectrum with source elimination} & &  \\ 
$+ 3 $ & $1.447^{+ 0.017}_{- 0.017}$ (1.4 fix) & $5.51^{+ 0.08}_{- 0.08}$ ($5.71^{+ 0.04}_{- 0.04}$) & 
$0.32^{+ 0.05}_{- 0.05}$ ($0.39^{+ 0.04}_{- 0.04}$) & 75/65 ~(95/66)\\ 
   0   & $1.418^{+ 0.017}_{- 0.017}$ (1.4 fix) & $5.70^{+ 0.08}_{- 0.08}$ ($5.78^{+ 0.04}_{- 0.04}$) & 
$0.36^{+ 0.04}_{- 0.05}$ ($0.39^{+ 0.04}_{- 0.04}$) & 72/65 ~(75/66)  \\
$- 3 $ & $1.392^{+ 0.016}_{- 0.017}$ (1.4 fix) & $5.88^{+ 0.08}_{- 0.08}$ ($5.85^{+ 0.04}_{- 0.04}$) & 
$0.40^{+ 0.05}_{- 0.04}$ ($0.39^{+ 0.04}_{- 0.04}$) & 78/65 ~(78/66)     \\ \hline 
\multicolumn{3}{l}{LSS spectrum without source elimination} & &  \\ 
$+ 3 $ & $1.469^{+ 0.015}_{- 0.015}$ (1.4 fix) & $5.97^{+ 0.08}_{- 0.07}$ ($6.29^{+ 0.04}_{- 0.04}$) & 
$0.42^{+ 0.04}_{- 0.05}$ ($0.54^{+ 0.04}_{- 0.03}$) & 75/65 (135/66)  \\ 
   0   & $1.442^{+ 0.015}_{- 0.014}$ (1.4 fix) & $6.17^{+ 0.08}_{- 0.07}$ ($6.36^{+ 0.04}_{- 0.04}$) & 
$0.46^{+ 0.05}_{- 0.04}$ ($0.54^{+ 0.03}_{- 0.04}$) & 70/65 ~(93/66)  \\
$- 3 $ & $1.417^{+ 0.015}_{- 0.014}$ (1.4 fix) & $6.35^{+ 0.08}_{- 0.08}$ ($6.43^{+ 0.04}_{- 0.04}$) & 
$0.50^{+ 0.05}_{- 0.04}$ ($0.53^{+ 0.04}_{- 0.03}$) & 75/65 ~(80/66)  \\ \hline
\end{tabular}
\end{center}
Note. Values in parentheses are results when $\Gamma^{\rm hard}$ is fixed to 1.4.
Errors are 90\% confidence levels.
Systematic errors are also examined by changing the NXB levels.
\end{table*}

\begin{table*}[htb]
\caption{Simulation results on the fluctuation width compared with
the analytical formula.}
\label{tab:lognlogs}
\begin{center}
\begin{tabular}{ccccccr}\hline
$k$ &  $\gamma$   & $S_{\rm min}$ (\ergs) & $N(>S_0)$ (deg$^{-2}$) & 
$\sigma_{\rm F}$ & Fluctuation & $N_{\rm F}$ \\ \hline\hline
$1.98\times 10^{-14}$ & 2.42 & $1.77\times 10^{-15}$ & 3.99 & 5.04~\% & $4.78^{+0.37}_{-0.40}$~\% & 83 \\
$1.58\times 10^{-15}$ & 2.50 & $2.52\times 10^{-15}$ & 3.59 & 5.06~\% & $5.06^{+0.19}_{-0.20}$~\% & 369 \\
$1.00\times 10^{-19}$ & 2.82 & $6.10\times 10^{-15}$ & 2.44 & 5.26~\% & $5.97^{+0.45}_{-0.49}$~\% & 79 \\
\hline
\end{tabular}
\end{center}
Note. Errors in the fluctuation are 1~$\sigma$.
\end{table*}

\begin{table*}[htb]
\caption{Simulation results on $\Gamma^{\rm hard}$
compared with the observation.}
\label{tab:SimGamma}
\begin{center}
\begin{tabular}{cccr}\hline
%% \\[-3.5ex]
$\Gamma_{\rm S}$  & ${\rm Av}(\Gamma^{\rm hard})$ & $\sqrt{{\rm Sd}^2(\Gamma^{\rm hard})}$ & $N_{\rm F}$
\\[0.5ex]
\hline\hline
$1.4\pm 0.0$ & $1.414\pm 0.001$ & $0.009^{+0.001}_{-0.001}$ & 369 \\
$1.2\pm 0.5$ & $1.397\pm 0.001$ & $0.020^{+0.001}_{-0.001}$ & 369 \\
$1.1\pm 1.0$ & $1.422\pm 0.003$ & $0.031^{+0.003}_{-0.003}$ & 249 \\
\hline
Observation  & $1.412\pm 0.007$ & $0.055^{+0.005}_{-0.006}$ & 91 \\
\hline
\end{tabular}
\end{center}
Note. Errors are 1~$\sigma$.
\end{table*}

\clearpage

\begin{figure}[htb] 
\begin{center}
\FigureFile(0.24\textwidth,0.5\textwidth){figure10a.eps}
\FigureFile(0.24\textwidth,0.5\textwidth){figure10b.eps}
\end{center}
\caption{Integrated pulse-height spectrum of the CXB for an exposure
time of 4.2~Ms and a sky coverage of 40~deg$^2$. Each panel shows
the data fitted with a 2-component model, the residual in unit of
$\sigma$, and the data to model ratio. The left panel is for the nominal
NXB subtraction, and the right one uses $3\%$ reduced NXB\@. 
%% The fitted parameters are indicated in each panel.
}\label{fig:2-10keV}
\end{figure}

\clearpage

\end{document}